\pdfoutput=1
\RequirePackage{ifpdf}
\ifpdf % We are running pdfTeX in pdf mode
\documentclass[pdftex]{sigma}
\else
\documentclass{sigma}
\fi

\usepackage{subfigure}

\begin{document}

\allowdisplaybreaks

\renewcommand{\thefootnote}{$\star$}

\renewcommand{\PaperNumber}{031}

\FirstPageHeading

\ShortArticleName{The Pascal Triangle of a~Discrete Image}

\ArticleName{The Pascal Triangle of a~Discrete Image:  Def\/inition, Properties and Application to Shape
Analysis\footnote{This paper is a~contribution to the Special Issue  ``Symmetries of Dif\/ferential
Equations:  Frames, Invariants and~Applications''.
The full collection is available at
\href{http://www.emis.de/journals/SIGMA/SDE2012.html}{http://www.emis.de/journals/SIGMA/SDE2012.html}}}

\Author{Mireille BOUTIN~$^\dag$ and Shanshan HUANG~$^\ddag$}

\AuthorNameForHeading{M.~Boutin and S.~Huang}

\Address{$^\dag$~School of Electrical and Computer Engineering, Purdue University, USA}
\EmailD{\href{mailto: mboutin@purdue.edu}{mboutin@purdue.edu}}

\Address{$^\ddag$~Department of Mathematics, Purdue University, USA}
\EmailD{\href{mailto: huang94@purdue.edu}{huang94@purdue.edu}}

\ArticleDates{Received September 24, 2012, in f\/inal form April 03, 2013; Published online April 11, 2013}

\Abstract{We def\/ine the Pascal triangle of a~discrete (gray scale) image as a~pyramidal arrangement of
complex-valued moments and we explore its geometric signif\/icance.
In particular, we show that the entries of row $k$ of this triangle correspond to the Fourier series
coef\/f\/icients of the moment of order $k$ of the Radon transform of the image.
Group actions on the plane can be naturally prolonged onto the entries of the Pascal triangle.
We study the prolongation of some common group actions, such as rotations and ref\/lections, and we propose
simple tests for detecting equivalences and self-equivalences under these group actions.
The motivating application of this work is the problem of characterizing the geometry of objects on images,
for example by detecting approximate symmetries.}


\Classification{30E05; 57S25; 68T10}

\renewcommand{\thefootnote}{\arabic{footnote}}
\setcounter{footnote}{0}

\section{Definition and reconstruction properties}

Let $\{(x_k, y_k)\}_{k=1}^N$ with $x_k, y_k\in\mathbb{R}$ represent the pixel locations of a~digital image.
For simplicity, we use complex coordinates $z_k=x_k+iy_k$.
Consider a~gray scale image def\/ined on $\{z_k\}_{k=1}^N$.
More specif\/ically, we have a~mapping $\rho:  \{z_k\}_{k=1}^N \rightarrow \mathbb{R}_{\geq0}$, where
$\rho(z_k)$ represents the intensity of pixel $z_k$.\footnote{We use $\mathbb{R}_{\geq0}$ instead of
a~specif\/ic discrete domain such as $\{0,1, \ldots,255\}$ for more generality.}
Denote the discrete gray scale image by $I=\{(z_k, \rho(z_k))\}_{k=1}^N$.

Consider the following moment matrix:
\begin{gather*}
\tau_N(I)=
\begin{pmatrix}
\mu_{0,0}&\mu_{1,0}&\mu_{2,0}&\mu_{3,0}&\cdots&\mu_{N-1,0}
\\
\mu_{0,1}&\mu_{1,1}&\mu_{2,1}&\cdots&\cdots&\mu_{N-1,1}
\\
\mu_{0,2}&\mu_{1,2}&\cdots&\cdots&\cdots&\mu_{N-1,2}
\\
\mu_{0,3}&\cdots&\cdots&\cdots&\cdots&\mu_{N-1,3}
\\
\vdots&\;&\;&\;&\;&\vdots
\\
\mu_{0, N-1}&\mu_{1, N-1}&\cdots&\cdots&\cdots&\mu_{N-1, N-1}
\end{pmatrix}
_{N\times N},
\end{gather*}
where
$
\mu_{j, l}=\sum\limits_{k=1}^N z_k^j\bar{z}_k^l\rho(z_k)$,
$
j, l\in\mathbb{Z}_{\geq0}$
is the complex moment of order $(j, l)$ for the discrete image~$I$.
Observe the conjugate symmetry property of the moments $\mu_{j, l}=\bar{\mu}_{l, j}$.
In particular, $\mu_{j, j}\in\mathbb{R}$, $\forall\,  j\in\mathbb{Z}_{\geq0}$.

We can express the relationship between the moments and the image $I$ in matrix form:
\begin{gather}
\label{tauZ}
\tau_N(I)=Z^\dag WZ,
\end{gather}
where
\begin{gather*}
Z=
\begin{pmatrix}
1&z_1&z_1^2&\cdots&z_1^{N-1}
\\
1&z_2&z_2^2&\cdots&z_2^{N-1}
\\
\vdots&\vdots&\cdots&\vdots
\\
1&z_N&z_N^2&\cdots&z_N^{N-1}
\end{pmatrix}
,
\qquad
W=
\begin{pmatrix}
\rho(z_1)&0&\cdots&0
\\
0&\rho(z_2)&\cdots&0
\\
\vdots&\vdots&\ddots&\vdots
\\
0&\cdots&0&\rho(z_N)
\end{pmatrix}
,
\end{gather*}
and $Z^\dag$ is the conjugate transpose of $Z$.
Observe that $Z$ is a~Vandermonde matrix and therefore is invertible when the pixel locations $z_k$ are
pairwise distinct.
Therefore, if the pixel coordinates are known and pairwise distinct, one can reconstruct the image $I$ by
matrix inversion:  $W=(Z^{-1})^\dag\tau_N(I)Z^{-1}$.
\begin{definition}
Let $r$ be a~nonnegative integer and let $I$ be a~discrete gray scale image.
The \textsl{Pascal triangle} $T^r(I)$ of order $r$ of $I$ is the following pyramid:
\begin{gather*}
\begin{array}
{@{}c@{\,}c@{\,}c@{\,}c@{\,}c@{\,}c@{\,}c@{\,}c@{\,}c@{\,}c@{\,}c@{\,}c@{\,}c@{}}&&&&&&\mu_{0,0}&&&&&&
\\
&&&&&\mu_{0,1}&&\mu_{1,0}&&&&&
\\
&&&&\mu_{0,2}&&2\mu_{1,1}&&\mu_{2,0}&&&&
\\
&&&\mu_{0,3}&&3\mu_{1,2}&&3\mu_{2,1}&&\mu_{3,0}&&&
\\
&&\mu_{0,4}&&4\mu_{1,3}&&6\mu_{2,2}&&4\mu_{3,1}&&\mu_{4,0}
\\[-0.95mm]
&&&&&&\vdots&&&&&&
\\[-0.95mm]
\mu_{0, r}&&\left(
\!\begin{smallmatrix}
r
\\
1
\end{smallmatrix}\!
\right)\mu_{1, r-1}&&\cdots&&\left(
\!\begin{smallmatrix}
r
\\
l
\end{smallmatrix}\!
\right)\mu_{l, r-l}&&\cdots&&\left(
\!\begin{smallmatrix}
r
\\
r-1
\end{smallmatrix}\!
\right)\mu_{r-1,1}&&\mu_{r,0}
\\
\end{array}
\end{gather*}
\end{definition}
\begin{lemma}[pixel intensity reconstruction property]\label{mutorholemma}
If the grid point locations $\{z_k\}_{k=1}^N$ are known and pairwise distinct, then the image
$I$ can be reconstructed from the Pascal triangle $T^{N-1}(I)$ of order $N-1$.
More specifically, knowledge of the entries of the right diagonal row of $T^{N-1}(I)$, i.e.\
$\{\mu_{j,0}\}_{j = 0}^{N-1}$, is sufficient for image reconstruction\footnote{The fact that the pixel intensities can be reconstructed
from a~f\/inite number of moments was stated in~\cite{Moment}.
Our lemma provides a~clear statement of the conditions under which this reconstruction is theoretically
possible.}.
\end{lemma}

\begin{proof}
Recall the def\/inition of the moments
\begin{gather*}
\mu_{j, l}=\sum_{k=1}^{N}z_k^j\bar{z}_k^l\rho(z_k).
\end{gather*}
We consider the vector formed by the moments $\{\mu_{j,0}\}_{j=0}^{N-1}$, which can be written in matrix form~as
\begin{gather}
\label{nmutorho}
\begin{pmatrix}
\mu_{0,0}
\\
\mu_{1,0}
\\
\mu_{2,0}
\\
\vdots
\\
\mu_{N-1,0}
\end{pmatrix}
=
\begin{pmatrix}
1&1&\cdots&1
\\
z_1&z_2&\cdots&z_N
\\
z_1^2&z_2^2&\cdots&z_N^2
\\
\vdots&\vdots&\vdots&\vdots
\\
z_1^{N-1}&z_2^{N-1}&\cdots&z_N^{N-1}
\end{pmatrix}
\begin{pmatrix}
\rho(z_1)
\\[2mm]
\rho(z_2)
\\[2mm]
\vdots
\\[2mm]
\rho(z_N)
\end{pmatrix}.
\end{gather}

Observe that the coef\/f\/icient matrix in~\eqref{nmutorho} is a~Vandermonde matrix.
The Vandermonde matrix has full rank when $z_j\neq z_k$ for all distinct $j, k = 1,2, \dots, N$.
Thus, since the pixel locations are assumed to be distinct, we can reconstruct the pixel
intensities $\{\rho(z_k)\}_{k=1}^N$ by inverting the coef\/f\/icient matrix and multiplying by
the moment vector on the left-hand-side.
\end{proof}

Notice that if we consider the Pascal triangle $T^{N}(I)$ of order $N$, then knowledge of the second right
diagonal row of $T^{N}(I)$, i.e.\
$\{\mu_{j,1}\}_{j = 0}^{N-1}$, is also suf\/f\/icient for image reconstruction as long as
the $z_k$'s are pairwise distinct and nonzero.
This is because the vector formed by the moments $\{\mu_{j,1}\}_{j=0}^{N-1}$ can be written in matrix form as
\begin{gather}
\begin{pmatrix}
\mu_{0,1}
\\
\mu_{1,1}
\\
\mu_{2,1}
\\
\vdots
\\
\mu_{N-1,1}
\end{pmatrix}
=
\begin{pmatrix}
\bar{z}_1&\bar{z}_2&\cdots&\bar{z}_N
\\
z_1\bar{z}_1&z_2\bar{z}_2&\cdots&z_N\bar{z}_N
\\
z_1^2\bar{z}_1&z_2^2\bar{z}_2&\cdots&z_N^2\bar{z}_N
\\
\vdots&\vdots&\vdots&\vdots
\\
z_1^{N-1}\bar{z}_1&z_2^{N-1}\bar{z}_2&\cdots&z_N^{N-1}\bar{z}_N
\end{pmatrix}
\begin{pmatrix}
\rho(z_1)
\\
\rho(z_2)
\\
\vdots
\\
\rho(z_N)
\end{pmatrix}
\nonumber\\
\hphantom{\begin{pmatrix}
\mu_{0,1}
\\
\mu_{1,1}
\\
\mu_{2,1}
\\
\vdots
\\
\mu_{N-1,1}
\end{pmatrix}}{}
=
\begin{pmatrix}
1&1&\cdots&1
\\
z_1&z_2&\cdots&z_N
\\
z_1^2&z_2^2&\cdots&z_N^2
\\
\vdots&\vdots&\vdots&\vdots
\\
z_1^{N-1}&z_2^{N-1}&\cdots&z_N^{N-1}
\end{pmatrix}
\begin{pmatrix}
\bar{z}_1&0&\cdots&0
\\
0&\bar{z}_2&\cdots&0
\\
\vdots&\vdots&\ddots&\vdots
\\
0&0&\cdots&\bar{z}_N
\end{pmatrix}
\begin{pmatrix}
\rho(z_1)
\\
\rho(z_2)
\\
\vdots
\\
\rho(z_N)
\end{pmatrix}.\label{nmutorho21}
\end{gather}
The coef\/f\/icient matrix in~\eqref{nmutorho21} is a~Vandermonde matrix multiplied by a~diagonal matrix.
Assuming that the pixel locations are pairwise distinct insures that the Vandermonde matrix is
invertible, and further assuming that they are nonzero insures invertibility of the diagonal matrix.
Hence the coef\/f\/icient matrix in~\eqref{nmutorho21} is nonsingular and we can reconstruct the pixel
intensities $\{\rho(z_k)\}_{k=1}^N $ from $ T^{N}(I) $ by inverting this coef\/f\/icient matrix and
multiplying by the moment vector on the left-hand-side.

A similar argument can be used to show that, for any f\/ixed $l$, the pixel intensities can be reconstructed
from the moment vector $\{\mu_{j, l}\}_{j = 0}^{N-1}$, which can be obtained from the Pascal
triangle $ T^{N+l-1}(I) $ of order $ N+l-1$.

\begin{remark}
 In practice, when reconstructing the pixel intensities of an image $ I$, f\/loating point errors in the
matrix inversion can result in inaccuracies in the reconstructed image.
In fact, the recovered pixel intensities may be complex valued.
While the imaginary part of the result tends to be quite small, it is advantageous to f\/irst reformulate
the problem to guarantee a~real solution.
One way to force the solution to be real is to separate equation~\eqref{nmutorho} into two sets of equations
with real coef\/f\/icients.
More specif\/ically, we can separate the equation system into its real part and its imaginary part, and
combine these two real equation systems into one.
After this, a~real solution for the new equation system can be found, for example, by singular value
decomposition (SVD).
\end{remark}

\begin{lemma}
Given the moments matrix $ \tau_N(I) $ of
a~discrete image $ I $ and an upper bound on the number $ N $ of pixels, one can reconstruct the pixel
location~$ z_k $ and the intensity~$ \rho(z_k) $ for all~$ z_k $ such that $ \rho(z_k) \neq 0$.\footnote{This result generalizes Proposition~1 in~\cite{Milanfar95}, which states that the vertices of
a~polygon are uniquely determined by a~f\/inite number of moments.}
\end{lemma}

\begin{proof}
If the number of pixels in the image $I $  is strictly less than $N$,  we can extend $I $  to an image with $N $
pixels by adding zero intensity pixels.
Without loss of generality, we assume that $\rho(z_k)\neq0 $  for $k=1, \dots, s $  and $\rho(z_k)=0 $  for
 $ k=s+1, \dots, N$.
Consider the polynomial
\begin{gather*}
P(t)=\prod_{k=1}^s(t-z_k)=t^s+\sum_{j=1}^s c_jt^{s-j},
\end{gather*}
where the coef\/f\/icients $c_j $  are polynomials in the $z_k$'s.

Observe that $P(z_k)=0$, $\forall\,  k=1,2, \dots, s$.  Therefore, we also have $\rho(z_k)\bar{z}_k^lP(z_k)=0 $,
for any $l=0, \dots, s-1$.  Summing all these equations over $k$'s, we get
\begin{gather*}
\sum_{k=1}^s\rho(z_k)\bar{z}_k^lP(z_k)=0 \quad \Longrightarrow
\quad
\sum_{k=1}^s\rho(z_k)\bar{z}_k^l\left(z_k^s+\sum_{j=1}^s c_jz_k^{s-j}\right)=0
\\
\hphantom{\sum_{k=1}^s\rho(z_k)\bar{z}_k^lP(z_k)=0} \quad \Longrightarrow
\quad
\sum_{k=1}^s\rho(z_k)\bar{z}_k^lz_k^s+\sum_{j=1}^s c_j\sum_{k=1}^s\rho(z_k)\bar{z}
_k^lz_k^{s-j}=0
\\
\hphantom{\sum_{k=1}^s\rho(z_k)\bar{z}_k^lP(z_k)=0} \quad \Longrightarrow
\quad
\mu_{s, l}+\sum_{j=1}^s c_j\mu_{s-j, l}=0
\\
\hphantom{\sum_{k=1}^s\rho(z_k)\bar{z}_k^lP(z_k)=0} \quad \Longrightarrow
\quad
\sum_{j=1}^s c_j\mu_{s-j, l}=-\mu_{s, l},
\qquad
l=0,1, \dots, s-1.
\end{gather*}
We write these last equations in matrix form:
\begin{gather}
\label{mutozk}
\underbrace{
\begin{pmatrix}
\mu_{0,0}&\mu_{1,0}&\mu_{2,0}&\mu_{3,0}&\cdots&\mu_{s-1,0}
\\
\mu_{0,1}&\mu_{1,1}&\mu_{2,1}&\cdots&\cdots&\mu_{s-1,1}
\\
\mu_{0,2}&\mu_{1,2}&\cdots&\cdots&\cdots&\mu_{s-1,2}
\\
\mu_{0,3}&\cdots&\cdots&\cdots&\cdots&\mu_{s-1,3}
\\
\vdots&\;&\;&\;&\;&\vdots
\\
\mu_{0, s-1}&\mu_{1, s-1}&\cdots&\cdots&\cdots&\mu_{s-1, s-1}
\end{pmatrix}
}
\begin{pmatrix}
c_s
\\
c_{s-1}
\\
c_{s-2}
\\
\vdots
\\
c_1
\end{pmatrix}
=-
\begin{pmatrix}
\mu_{s,0}
\\
\mu_{s,1}
\\
\mu_{s,2}
\\
\vdots
\\
\mu_{s, s-1}
\end{pmatrix}
.
\\
\hspace{35mm}\tau_s(I)\nonumber
\end{gather}

From equation~\eqref{tauZ} we know that
\begin{gather*}
\tau_s(I)=
\begin{pmatrix}
1&1&\cdots&1
\\
\bar{z}_1&\bar{z}_2&\cdots&\bar{z}_s
\\
\vdots&\cdots&\vdots
\\
\bar{z}_1^{s-1}&\bar{z}_2^{s-1}&\cdots&\bar{z}_s^{s-1}
\end{pmatrix}\!
\begin{pmatrix}
\rho(z_1)&0&\cdots&0
\\
0&\rho(z_2)&\cdots&0
\\
\vdots&\vdots&\ddots&\vdots
\\
0&\cdots&0&\rho(z_s)
\end{pmatrix}\!
\begin{pmatrix}
1&z_1&z_1^2&\cdots&z_1^{s-1}
\\
1&z_2&z_2^2&\cdots&z_2^{s-1}
\\
\vdots&\vdots&\cdots&\vdots
\\
1&z_s&z_s^2&\cdots&z_s^{s-1}
\end{pmatrix}\!
.
\end{gather*}
Thus $\tau_s(I) $  is invertible, since the locations $z_k $  are pairwise distinct and the pixel intensi\-ties~ $ \rho(z_k) $  are nonzero.
Hence we can solve the above equation system for $(c_s, c_{s-1}, \dots, c_1) $  by inver\-ting~$\tau_s(I) $  and
multiplying by the vector on the right-hand-side of equation~\eqref{mutozk}.

Since the $c_k$'s determine the polynomial $P(t)$,  we can solve for the roots of $P(t)=0$,  which are
actually $\{z_k\}_{k=1}^s $.
By Lemma~\ref{mutorholemma}, we can subsequently obtain the pixel intensities
 $\{\rho(z_k)\}_{k=1}^s $.
\end{proof}

\begin{remark}
To determine the number of nonzero pixels, we can look at the rank of $\tau_N(I)$.  Since
 $ \tau_N(I)=Z^\dag WZ $  by equation~\eqref{tauZ} and $\text{rank}(Z^\dag)=\text{rank}(Z)=N $,
 $ \text{rank}(W)=s$,  we can conclude that $\text{rank}(\tau_N(I))=s$.
 \end{remark}

Since the Pascal triangle $T^{2N-2}(I) $  of the image $I $  contains all the information needed to recover
 $ \tau_N(I)$,  we have the following corollary:
\begin{corollary}[image reconstruction property]
\label{Ttoimage}
Given the Pascal triangle $T^{2N-2}(I) $  of a~discrete image $I$,  one can reconstruct both the grid point
locations $\{z_k\}_{k=1}^N $  and the corresponding intensities
 $\{\rho(z_k)\}_{k=1}^N $  for all those $z_k $  such that $\rho(z_k)\neq0$.
\end{corollary}

\section{Relationship with the Radon transform}
The Radon transform $f_{\theta}(r)$  is the projection of
the image $I=\{(z_k, \rho(z_k))\}_{k=1}^N $  onto the straight line through the origin
with direction vector $
\begin{pmatrix}
\cos(\theta) & \sin(\theta)
\end{pmatrix}
^T$, i.e.\
\begin{gather*}
f_{\theta}(r)=\sum_{k\in S}\rho(z_k),
\end{gather*}
where $ S = \{k\,|\, x_k\cos(\theta)+y_k\sin(\theta) = r, \ k = 1,2, \dots, N\}$.  Since $ f_{\theta}(r) $ is a~periodic
function of~$ \theta $ with period~$ 2\pi$, any of its $n$-th order moment $ m_n(\theta) $ is also periodic with
period $ 2\pi$.
It turns out that, for any $ n=0,1,2, \ldots$, the coef\/f\/icients of the Fourier series of $ m_n(\theta) $ are
given by the entries of row $ (n+1) $ of $ T^r(I) $ with $ r \geq n$.
\begin{lemma}
%\label{mutomn}
The $n$-th order moment $ m_n(\theta) $ of the Radon transform $ f_\theta(r) $ is given by the following linear
combination of the $ (n+1)$-th row entries of the Pascal triangle $ T^r(I) $ with $ r \geq n $:
\begin{gather}
\label{mnmukl}
m_n(\theta)=\frac{1}{2^n}\sum_{l=0}^n
\begin{pmatrix}
n
\\
l
\end{pmatrix}
\mu_{l, n-l}e^{i(n-2l)\theta}.
\end{gather}
\end{lemma}
\begin{proof}
For $n=0$, we have
\begin{gather*}
m_0(\theta)=\sum_{k=1}^N\rho(z_k)=\sum_{k=1}^N\rho(z_k)z_k^0\bar{z}_k^0=\mu_{0,0}.
\end{gather*}
For $n>0$, we have
\begin{gather*}
m_n(\theta)=\sum_r r^nf_\theta(r)=\sum_{k=1}^N\big(r_k(\theta)\big)^n\rho(z_k),
\end{gather*}
where $ r_k(\theta) $ is the projection of the vector $ (x_k, y_k)^T $ onto the axis with angle $ \theta \in
(-\pi, \pi] $ with respect to $x$-axis.
More precisely,
\begin{gather*}
r_k(\theta)=x_k\cos\theta+y_k\sin\theta=\frac{1}{2}\big(z_ke^{-i\theta}+\bar{z}_ke^{i\theta}\big),
\qquad
\forall\,  k=1,2, \dots, N,
\end{gather*}
and therefore
\begin{gather*}
m_n(\theta)
=\sum_{k=1}^N\big(r_k(\theta)\big)^n\rho(z_k)
=\sum_{k=1}^N\left(\frac{1}{2}\big(z_ke^{-i\theta}+\bar{z}_ke^{i\theta}\big)\right)^n\rho(z_k)
\\
\phantom{m_n(\theta)}
{}=\frac{1}{2^n}\sum_{k=1}^N\big(z_ke^{-i\theta}+\bar{z}_ke^{i\theta}\big)^n\rho(z_k)
=\frac{1}{2^n}\sum_{k=1}^N\left(\sum_{l=0}^n
\begin{pmatrix}
n\\l
\end{pmatrix}
z_k^le^{-il\theta}\bar{z}_k^{n-l}e^{i(n-l)\theta}\right)\rho(z_k)
\\
\phantom{m_n(\theta)}
{}=\frac{1}{2^n}\sum_{l=0}^n
\begin{pmatrix}
n\\l
\end{pmatrix}
\left(\sum_{k=1}^Nz_k^l\bar{z}_k^{n-l}\rho(z_k)\right)e^{-il\theta}e^{i(n-l)\theta}
=\frac{1}{2^n}\sum_{l=0}^n
\begin{pmatrix}
n\\l
\end{pmatrix}
\mu_{l, n-l}e^{i(n-2l)\theta}.\tag*{\qed}
\end{gather*}
  \renewcommand{\qed}{}
\end{proof}

\begin{figure}[t]
  \centering
\includegraphics[scale=0.9]{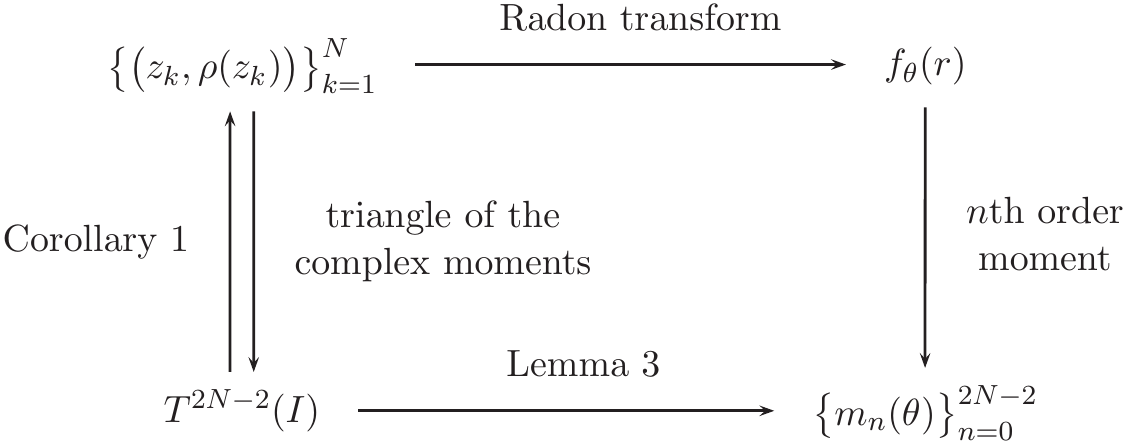}
  \caption{Relationship between the Pascal triangle and the Radon transform of an image under the assumption that the pixel locations are unknown a priori.}\label{F1}
\end{figure}

\begin{figure}[t]
  \centering
  \includegraphics[scale=0.9]{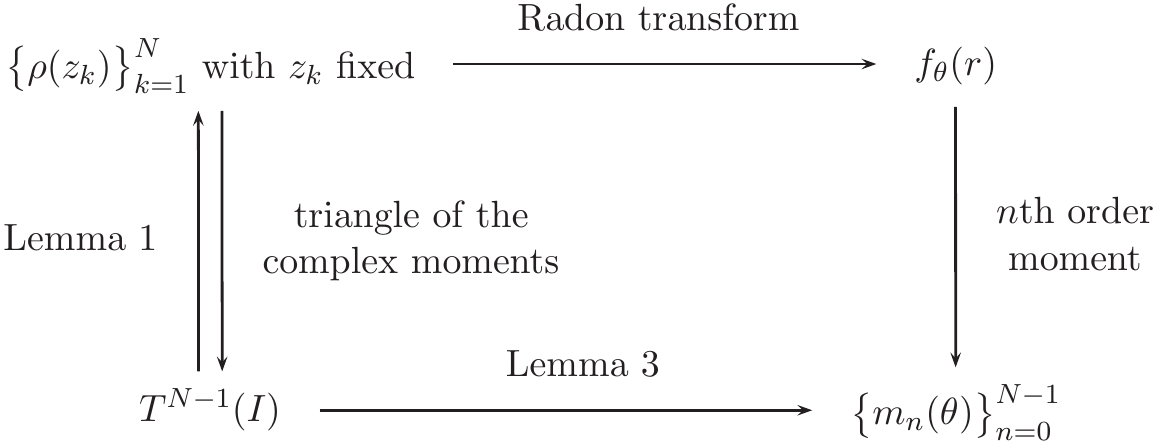}
  \caption{Relationship between the Pascal triangle and the Radon transform of an image under the assumption that the pixel locations  are fixed and known a priori.}\label{F2}
\end{figure}

Figs.~\ref{F1} and~\ref{F2} summarize the relationship between the Pascal triangle and the Radon
transform of an image when the pixel locations are known and unknown, respectively.
Observe that a~smaller number of rows of the Pascal triangle are needed in order to reconstruct the image
if the pixel locations were known.

\section{Image reconstruction from samples}
\begin{lemma}
\label{mtoTn}
The last row of $ T^n(I) $ can be reconstructed from $ n+1 $ generic moment samples $ m_n(\theta_1), m_n(\theta_2), \dots, m_n(\theta_{n+1})$.
\end{lemma}
\begin{proof}
We f\/irst write equation~\eqref{mnmukl} in matrix form:
\begin{gather*}
\begin{pmatrix}
m_n(\theta_1)
\\
m_n(\theta_2)
\\
\vdots
\\
m_n(\theta_{n+1})
\end{pmatrix}
=\frac{1}{2^n}
\begin{pmatrix}
e^{in\theta_1}&\cdots&e^{i(n-2l)\theta_1}&\cdots&e^{-in\theta_1}
\\
e^{in\theta_2}&\cdots&e^{i(n-2l)\theta_2}&\cdots&e^{-in\theta_2}
\\
\vdots&\cdots&\vdots&\cdots&\vdots
\\
e^{in\theta_{n+1}}&\cdots&e^{i(n-2l)\theta_{n+1}}&\cdots&e^{-in\theta_{n+1}}
\end{pmatrix}
\begin{pmatrix}
\mu_{0, n}
\\
\vdots
\\
\left(
\begin{smallmatrix}
n
\\
l
\end{smallmatrix}
\right)\mu_{l, n-l}
\\
\vdots
\\
\mu_{n,0}
\end{pmatrix}
.
\end{gather*}
There is a~unique solution for $\big(\mu_{0, n}, \dots, \left(
\begin{smallmatrix}
n
\\
l
\end{smallmatrix}
\right)\mu_{l, n-l}, \dots, \mu_{n,0}\big) $ if and only if
\begin{gather*}
\det\left[
\begin{pmatrix}
e^{in\theta_1}&\cdots&e^{i(n-2l)\theta_1}&\cdots&e^{-in\theta_1}
\\
e^{in\theta_2}&\cdots&e^{i(n-2l)\theta_2}&\cdots&e^{-in\theta_2}
\\
\vdots&\cdots&\vdots&\cdots&\vdots
\\
e^{in\theta_{n+1}}&\cdots&e^{i(n-2l)\theta_{n+1}}&\cdots&e^{-in\theta_{n+1}}
\end{pmatrix}
\right]\neq0
\\
\Longleftrightarrow
\
\det\left[ %\text~2
\begin{pmatrix}
\!e^{-in\theta_1}\!&0&0&\dots&0
\\
0&\!e^{-in\theta_2}\!&0&\dots&0
\\
\vdots&\vdots&\ddots&&\vdots
\\
0&\dots&0&\dots&\!e^{-in\theta_{n+1}}\!
\end{pmatrix}\!
\begin{pmatrix}
\!e^{i2n\theta_1}\!&\cdots&\!e^{i4\theta_1}\!&\!e^{i2\theta_1}\!&1
\\
\!e^{i2n\theta_2}\!&\cdots&\!e^{i4\theta_2}\!&\!e^{i2\theta_2}\!&1
\\
\vdots&\cdots&\vdots&\vdots&\vdots
\\
\!e^{i2n\theta_{n+1}}\!&\cdots&\!e^{i4\theta_{n+1}}\!&\!e^{i2\theta_{n+1}}\!&1
\end{pmatrix}
\right]
\!\neq0
\\
\Longleftrightarrow
\
\det\left[ %\text~2
\begin{pmatrix}
e^{i2n\theta_1}&\cdots&e^{i4\theta_1}&e^{i2\theta_1}&1
\\
e^{i2n\theta_2}&\cdots&e^{i4\theta_2}&e^{i2\theta_2}&1
\\
\vdots&\cdots&\vdots&\vdots&\vdots
\\
e^{i2n\theta_{n+1}}&\cdots&e^{i4\theta_{n+1}}&e^{i2\theta_{n+1}}&1
\end{pmatrix}
\right] \neq0,
\end{gather*}
since $e^{-in\theta_j}\neq0$, $\forall\,  j=1, \dots, n + 1$.

Observe that the above determinant is a~Vandermonde determinant.
It is nonzero if $ e^{i2\theta_j} \neq e^{i2\theta_k} $ for all distinct $ j, k = 1, \dots, n+1 $.
Therefore, if the $ \theta_j$'s are such that $ e^{i2\theta_j} \neq e^{i2\theta_k} $ (thus the need to pick
a~generic sample set), we will get a~unique solution for $\big(\mu_{0, n}, \dots, \left(
\begin{smallmatrix}
n
\\
l
\end{smallmatrix}
\right)\mu_{l, n-l}, \dots, \mu_{n,0}\big) $.
Hence we can reconstruct the last row of $ T^n(I) $.
\end{proof}

\begin{corollary}%\label{sptorho}
Given the grid point
locations $\{z_k\}_{k=1}^N$, we can reconstruct the discrete image $ I =
\{(z_k, \rho(z_k))\}_{k=1}^N $ from the Radon transform $ f_{\theta_1}(r),
f_{\theta_2}(r), \dots, f_{\theta_N}(r) $ at $ N $ fixed generic angles $ \theta_1, \dots, \theta_N $.\footnote{This result generalizes Theorem~5.1 in~\cite{Milanfar00}, which states that a~quadrature domain can
be uniquely reconstructed by the line integral projections at f\/inite angles.}
\end{corollary}
\begin{proof}
From the given radon transform of the image at dif\/ferent angles, we can calculate the
moments $ \big\{m_n(\theta_k)\, |\, n = 0, \dots, N-1, \; k = 1, \dots, n+1\big\}$.  The conclusion followed by
combining Lemmas~\ref{mutorholemma} and~\ref{mtoTn}.
\end{proof}

\begin{figure}[h!]
\includegraphics[scale=0.85]{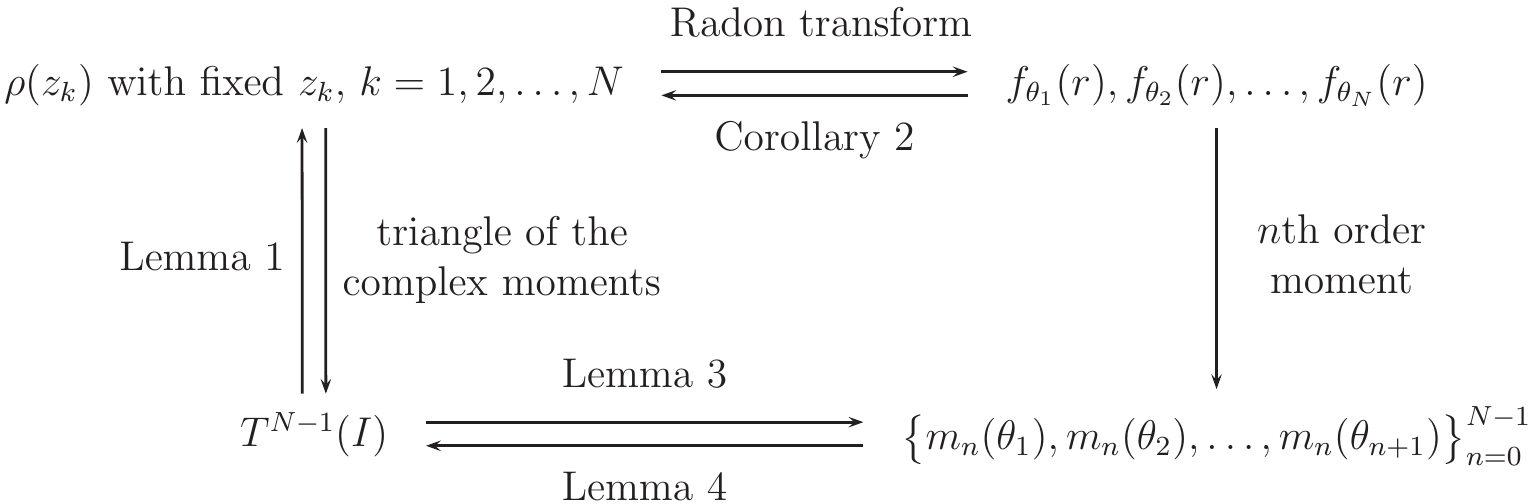}
\centering\caption{Relationship between the Pascal triangle $ T^{N-1}(I) $ and the Radon transform.}
\label{F3}
\end{figure}

The diagram of Fig.~\ref{F3} thus commutes.
Note that, one could use a~similar argument along with Corollary~\ref{Ttoimage} to show that $(2N-1) $  generic
observations of $ \{m_n(\theta_j),\, j = 1, \ldots, n+1 \}_{n=0}^{2N-2}$ would be needed to fully
reconstruct the image $ \{ (z_k, \rho(z_k) ) \}_{k=1}^N $  with pixel
positions $z_k $  unknown.

\section{Prolongation of group actions on the moments\\ and invariantization}
\label{sec4}

Let $ (G, \cdot) $ be a~group acting on the complex plane:
\begin{gather*}
\cdot: \  G\times\mathbb{C}\longrightarrow\mathbb{C},\\
\hspace*{7.4mm} (g, z) \longmapsto  g\cdot z,
\qquad
\forall\,  g\in G, \ z\in\mathbb{C}.
\end{gather*}
This induces a~group transformation $ (G, \circ) $ of the discrete image $I= \{ (z_k, \rho(z_k) ) \}_{k=1}^N$, namely
\begin{gather*}
g\circ \{ (z_k, \rho(z_k) ) \}_{k=1}^N= \{ (g\cdot z_k, \rho(z_k) ) \}_{k=1}^N,
\qquad
\forall\,  g\in G.
\end{gather*}
Then the induced transformation $ (G, \ast) $ on moments $ \{\mu_{j, l}\}_{j, l\in\mathbb{Z}_\geq0} $ is{\samepage
\begin{gather*}
g\ast\{\mu_{j, l}\}_{j, l\in\mathbb{Z}_\geq0}=g\ast\left\{\sum_{k=1}^Nz_k^j\bar{z}_k^l\rho(z_k)\right\}_{j, l\in\mathbb{Z}_\geq0}
=\left\{\sum_{k=1}^N(g\cdot z_k)^j(\overline{g\cdot z_k})^l\rho(z_k)\right\}_{j, l\in\mathbb{Z}_\geq0}.
\end{gather*}
In other words, the transformed moments are the moments of the transformed image.}

\begin{example}
\label{exp1}
Consider the action of $ G = \mathbb{C} $ on $ \mathbb{C} $ by translation
\begin{gather*}
(z_0, z)\mapsto z+z_0, \qquad \forall\,  z_0\in G, \qquad \forall\,  z\in\mathbb{C}.
\end{gather*}
Then the induced transformation on the image $ I =  \{ (z_k, \rho(z_k) ) \}_{k=1}^N $ is
\begin{gather}
\label{example1}
z_0\circ \{ (z_k, \rho(z_k) ) \}_{k=1}^N= \{ (z_k+z_0, \rho(z_k) ) \}
_{k=1}^N,
\qquad
\forall\,  z_0\in G.
\end{gather}
In other words, the image is translated horizontally with distance $ x_0 = \text{Re}(z_0) $ and vertically
with $ y_0 = \text{Im}(z_0)$.
The transformed complex moments are
\begin{gather}
\tilde{\mu}_{j, l}
=\sum_{k=1}^N\tilde{z}_k^j\bar{\tilde{z}}_k^l\rho(\tilde{z}_k)=\sum_{k=1}
^N(z_k+z_0)^j(\bar{z}_k+\bar{z}_0)^l\rho(z_k)
\nonumber
\\
\phantom{\tilde{\mu}_{j, l}}
{}=\sum_{k=1}^N\left(\sum_{s=0}^j
\begin{pmatrix}j\\s\end{pmatrix}
z_k^sz_0^{j-s}\right)
\left(\sum_{t=0}^l
\begin{pmatrix}l\\t\end{pmatrix}
\bar{z}_k^t\bar{z}_0^{l-t}\right)\rho(z_k)
\nonumber
\\
\phantom{\tilde{\mu}_{j, l}}
{}=\sum_{s=0}^j\sum_{t=0}^l\left(\sum_{k=1}^Nz_k^s\bar{z}_k^t\rho(z_k)\right)
\begin{pmatrix}j\\s\end{pmatrix}
\begin{pmatrix}l\\t\end{pmatrix}
z_0^{j-s}\bar{z}_0^{l-t}
\nonumber
\\
\phantom{\tilde{\mu}_{j, l}}
{}=\sum_{s=0}^j\sum_{t=0}^l\mu_{s, t}
\begin{pmatrix}j\\s\end{pmatrix}
\begin{pmatrix}l\\t\end{pmatrix}
z_0^{j-s}\bar{z}_0^{l-t},
\qquad
\forall\,  j, l\in\mathbb{Z}_{\geq0}.
\label{transmu}
\end{gather}

Written in matrix form, the transformation of the moment matrix $ \tau_N(I) $ is $ \tau_N(\tilde{I}) = A^\dag
\tau_N(I)A$, where $ A = (a_{j, l})_{N\times N} $ is an upper-triangular matrix with $ a_{j, l} = \left(
\begin{smallmatrix}
l-1
\\
j-1
\end{smallmatrix}
\right)z_0^{l-j}$, and $ A^\dag $ is the conjugate transpose of $ A$.

\looseness=-1
Having obtained an explicit formula for the action of $ G $ on the moments, we follow Fels and Olver's moving
frame method~\cite{ MFI,MFII, Olverbook} to obtain a~set of invariant functions of the moments.
More specif\/ically, we consider the cross-section def\/ined by $ \tilde{\mu}_{1,0} = 0$.
The group transformation that maps $ \tau_N(I) $ to the cross-section is the moving frame $ z_0 =
-\frac{\mu_{1,0}}{\mu_{0,0}}$.
By applying the moving frame to the moment matrix, we obtain the matrix $ \tilde{\tau}_N(I) =
A_0^\dag\tau_N(I)A_0$, where $ A_0 = \left(\left(
\begin{smallmatrix}l-1\\j-1\end{smallmatrix}
\right)(-\frac{\mu_{1,0}}{\mu_{0,0}})^{l-j}\right)_{N\times N}$.  By equivariance of the moving frame, all
the entries of $ \tilde{\tau}_N(I) $ are invariant under translation.
One can check that theses entries $ \tilde{\mu}_{j, l} $ are actually the centralized moments
\begin{gather}
\label{centralmoments}
\tilde{\mu}_{j, l}=\sum_{k=1}^N\left(z_k-\frac{\mu_{1,0}}{\mu_{0,0}}\right)^j
\left(\bar{z}_k-\frac{\bar{\mu}_{1,0}
}{\mu_{0,0}}\right)^l\rho(z_k), \qquad j, l\in\mathbb{Z}_{\geq0}.
\end{gather}

By normalizing  (i.e.\
applying the moving frame transformation to) the coordinates of $ T^r(I)$, we obtain the translation invariant
Pascal triangle $ T_{\rm trans}^r(I) $ for a~discrete image $ I $:
\begin{gather*}
\begin{array}{@{}c@{\,}c@{\,}c@{\,}c@{\,}c@{\,}c@{\,}c@{\,}c@{\,}c@{\,}c@{\,}c@{\,}c@{\,}c@{}}&&&&&&\tilde{\mu}_{0,0}&&&&&&
\\
&&&&&\tilde{\mu}_{0,1}&&\tilde{\mu}_{1,0}&&&&&
\\
&&&&\tilde{\mu}_{0,2}&&2\tilde{\mu}_{1,1}&&\tilde{\mu}_{2,0}&&&&
\\
&&&\tilde{\mu}_{0,3}&&3\tilde{\mu}_{1,2}&&3\tilde{\mu}_{2,1}&&\tilde{\mu}_{3,0}&&&
\\
&&\tilde{\mu}_{0,4}&&4\tilde{\mu}_{1,3}&&6\tilde{\mu}_{2,2}&&4\tilde{\mu}_{3,1}&&\tilde{\mu}_{4,0}
\\
&&&&&&\vdots&&&&&&
\\
\tilde{\mu}_{0, r}&&\left(
\begin{smallmatrix}
r
\\
1
\end{smallmatrix}
\right)\tilde{\mu}_{1, r-1}&&\cdots&&\left(
\begin{smallmatrix}
r
\\
l
\end{smallmatrix}
\right)\tilde{\mu}_{l, r-l}&&\cdots&&\left(
\begin{smallmatrix}
r
\\
r-1
\end{smallmatrix}
\right)\tilde{\mu}_{r-1,1}&&\tilde{\mu}_{r,0}
\\
\end{array}
\end{gather*}

Observe that the corresponding $ n $-th order central moment $ \tilde{m}_n(\theta) $ of the image,
\begin{gather*}
\tilde{m}_n(\theta)=\sum_{k=1}^N\big(r_k(\theta)-r_0(\theta)\big)^n\rho(x_k, y_k),
\end{gather*}
where $ r_0(\theta)= x_0\cos\theta + y_0\sin\theta $ is the projection of the centroid, is invariant under
translations.
\end{example}

\begin{lemma}[orbit separation property of $ T^{2N-2}_{\rm trans}(I)$] Let $ I_1$, $I_2 $ be two discrete gray scale images with the
same number $N$ of pixels.
There exists a~translation $ g\in \mathbb{C} $ such that $ g\circ I_1 = I_2$, where $ \circ $ is defined as
in~\eqref{example1} $ \Longleftrightarrow $ $T^r_{\rm trans}(I_1) = T^r_{\rm trans}(I_2) $ for $r\geq 2N-2$.
\end{lemma}

\begin{proof}
 $ \Rightarrow $ If $ \exists\,g\in G $ such that $ g\circ I_1 = I_2$, we have $ z_k^{(2)} = z_k^{(1)} +
z_0 $ and $ \rho_2\big(z_k^{(2)}\big) = \rho_1\big(z_k^{(1)}\big)$, for some $ z_0 \in \mathbb{C}$, $k = 1, \dots, N$.
From equation~\eqref{transmu} we know that
\begin{gather*}
\mu_{0,0}^{(2)} = \mu_{0,0}^{(1)},
\qquad
 \mu_{1,0}^{(2)} = \mu_{0,0}^{(1)}z_0 + \mu_{1,0}^{(1)}, \qquad \text{hence} \qquad
\frac{\mu_{1,0}^{(2)}}{\mu_{0,0}^{(2)}}=\frac{\mu_{1,0}^{(1)}}{\mu_{0,0}^{(1)}}+z_0.
\end{gather*}

Then applying equation~\eqref{centralmoments}, we can get for any $ j, l\in\mathbb{Z}_{\geq 0}$
\begin{gather*}
\tilde{\mu}_{j, l}^{(1)}
=\sum_{k=1}^N\left(z_k^{(1)}-\frac{\mu_{1,0}^{(1)}}{\mu_{0,0}^{(1)}}\right)^j\left(\bar{z}
_k^{(1)}-\frac{\bar{\mu}_{1,0}^{(1)}}{\mu_{0,0}^{(1)}}\right)^l\rho_1\big(z_k^{(1)}\big)
\\
\phantom{\tilde{\mu}_{j, l}^{(1)}}
{}=\sum_{k=1}^N(z_k^{(2)}
-\frac{\mu_{1,0}^{(1)}}{\mu_{0,0}^{(1)}}-z_0)^j\left(\bar{z}_k^{(2)}-\frac{\bar{\mu}_{1,0}^{(1)}}{\mu_{0,0}^{(1)}
}-\bar{z}_0\right)^l\rho_2\big(z_k^{(2)}\big)=\tilde{\mu}_{j, l}^{(2)}.
\end{gather*}
Therefore $T^r_{\rm trans}(I_1)=T^r_{\rm trans}(I_2) $  for any $r\in\mathbb{Z}_{\geq0} $.

 $ \Leftarrow $ If $ T^r_{\rm trans}(I_1) = T^r_{\rm trans}(I_2) $ for $r\geq 2N-2$, from Corollary~\ref{Ttoimage}, we
conclude that $ I_1^{\rm trans} = I_2^{\rm trans}$, i.e.
\begin{gather*}
\big\{\big(z_{k, {\rm trans}}^{(1)}, \rho_1\big(z_{k, {\rm trans}}^{(1)}\big)\big)\big\}_{k=1}^N=\big\{\big(
z_{k, {\rm trans}}^{(2)}, \rho_2\big(z_{k, {\rm trans}}^{(2)}\big)\big)\big\}_{k=1}^N.
\end{gather*}
Hence $ \exists\, z_0, z_0'\in \mathbb{C} $ s.t.\
 $ z_{k, {\rm trans}}^{(1)} = z_k^{(1)} + z_0$, $z_{k, {\rm trans}}^{(2)} = z_k^{(2)} + z_0' $ with $ \rho_1\big(z_{k, {\rm trans}}^{(1)}\big)
= \rho_1\big(z_k^{(1)}\big) $ and $ \rho_2\big(z_{k, {\rm trans}}^{(2)}\big) = \rho_2\big(z_k^{(2)}\big) $ for any $ k = 1,2, \dots, N$.  Without
loss of generality, we assume that $ z_{k, {\rm trans}}^{(1)} = z_{k, {\rm trans}}^{(2)} $ and $ \rho_1\big(z_{k, {\rm trans}}^{(1)}\big) =
\rho_2\big(z_{k, {\rm trans}}^{(2)}\big)$.  Therefore $ \exists\, z_0 - z_0'\in \mathbb{C} = G $ satisfying
\begin{gather*}
z_k^{(1)}+z_0-z_0'=z_k^{(2)},
\qquad
\rho_1\big(z_k^{(1)}+(z_0-z_0')\big)=\rho_1\big(z_{k}^{(1)}\big)=\rho_2\big(z_k^{(2)}\big),
\end{gather*}
i.e.
 $ \exists\;g=z_0'-z_0\in G = \mathbb{C} $ such that $ g\circ I_1 = I_2$.
\end{proof}

\begin{remark} Without loss of generality, we can assume that the two images have the same number of pixels by
simply adding zero valued pixels to the smaller image.
\end{remark}

\begin{example}
\label{exp2}
Consider the action of $G=\mathbb{R}_+ $  on $\mathbb{C}$ by scaling
\begin{gather*}
(\lambda, z)\mapsto\lambda z, \qquad \forall\, \lambda\in G, \qquad \forall\,  z\in\mathbb{C}.
\end{gather*}
Then the induced transformation on the image $I= \{ (z_k, \rho(z_k) ) \}_{k=1}^N $  is
\begin{gather}\label{example2}
\lambda\circ \{ (z_k, \rho(z_k) ) \}_{k=1}^N= \{ (\lambda z_k, \rho(z_k) )
 \}_{k=1}^N,
\qquad
\forall\, \lambda\in G.
\end{gather}
In other words, the image is scaled by a~factor $\lambda $  both horizontally and vertically.
Then the transformed complex moments are
\begin{gather*}
\hat{\mu}_{j, l}=\sum_{k=1}^N\hat{z}_k^j\bar{\hat{z}}_k^l\rho(\hat{z}_k)=\sum_{k=1}
^N(\lambda z_k)^j(\lambda\bar{z}_k)^l\rho(z_k)
\\
\hphantom{\hat{\mu}_{j, l}}{} =\sum_{k=1}^N\lambda^{j+l}z_k^j\bar{z}_k^l\rho(z_k)=\lambda^{j+l}\sum_{k=1}^Nz_k^j\bar{z}_k^l\rho(z_k)
=\mu_{j, l}\lambda^{j+l},
\qquad
\forall\,  j, l\in\mathbb{Z}_{\geq0}.
\end{gather*}
Written in matrix form, the moment matrix for the new image $\hat{I}$ after scaling is
\begin{gather*}
\tau_N(\hat{I})=
\begin{pmatrix}
1&0&0&\cdots&0
\\
0&\lambda&0&\cdots&0
\\
0&0&\lambda^{2}&\cdots&0
\\
\vdots&\vdots&\vdots&\ddots&\vdots
\\
0&\cdots&\cdots&0&\lambda^{N-1}
\end{pmatrix}
\tau_N(I)
\begin{pmatrix}
1&0&0&\cdots&0
\\
0&\lambda&0&\cdots&0
\\
0&0&\lambda^{2}&\cdots&0
\\
\vdots&\vdots&\vdots&\ddots&\vdots
\\
0&\cdots&\cdots&0&\lambda^{N-1}
\end{pmatrix}
.
\end{gather*}

Again, we use the moving frame method of Fels and Olver to obtain a~set of invariant functions of the
moments.
Notice that
$\hat{\mu}_{1,1}=\sum\limits_{k=1}^N\hat{z}_k\bar{\hat{z}}_k\rho(\hat{z}_k)=\sum\limits_{k=1}^N\big(\hat{x}_k^2+\hat{y}_k^2\big)\rho(\hat{z}_k)\neq0 $
unless all $\rho(z_k) $  are zero.
We consider the cross-section def\/ined by $\hat{\mu}_{1,1}=1$.
The group transformation that maps $\tau_N(I) $  to the cross-section is the moving frame
 $ \lambda=(\mu_{1,1})^{-\frac{1}{2}}$.
By applying the moving frame to the moment matrix, we obtain the matrix
\begin{gather*}
\hat{\tau}_N(I)=
\begin{pmatrix}
1&0&\cdots&0
\\
0&(\mu_{1,1})^{-\frac{1}{2}}&\cdots&0
\\
\vdots&\vdots&\ddots&\vdots
\\
0&\cdots&0&(\mu_{1,1})^{-\frac{N-1}{2}}
\end{pmatrix}
\tau_N(I)
\begin{pmatrix}
1&0&\cdots&0
\\
0&(\mu_{1,1})^{-\frac{1}{2}}&\cdots&0
\\
\vdots&\vdots&\ddots&\vdots
\\
0&\cdots&0&(\mu_{1,1})^{-\frac{N-1}{2}}
\end{pmatrix}
.
\end{gather*}
By equivariance of the moving frame, all the entries of $\hat{\tau}_N(I) $  are invariant under scaling.

By normalizing the coordinates of $T^r(I)$,  we obtain the scaling invariant Pascal triangle
 $ T^r_{\rm scale}(I) $  for a~discrete image $I $:
\begin{gather*}
\begin{array}{@{}c@{\,}c@{\,}c@{\,}c@{\,}c@{\,}c@{\,}c@{\,}c@{\,}c@{\,}c@{\,}c@{\,}c@{\,}c@{}}&&&&&&\mu_{0,0}&&&&&&
\\
&&&&&\frac{\mu_{0,1}}{\sqrt{\mu_{1,1}}}&&\frac{\mu_{1,0}}{\sqrt{\mu_{1,1}}}&&&&&
\\
&&&&\frac{\mu_{0,2}}{\mu_{1,1}}&&2&&\frac{\mu_{2,0}}{\mu_{1,1}}&&&&
\\
&&&\frac{\mu_{0,3}}{\mu_{1,1}^{3/2}}&&3\frac{\mu_{1,2}}{\mu_{1,1}^{3/2}}&&3\frac{\mu_{2,1}}{\mu_{1,1}^{3/2}}
&&\frac{\mu_{3,0}}{\mu_{1,1}^{3/2}}&&&
\\
&&\frac{\mu_{0,4}}{\mu_{1,1}^2}&&4\frac{\mu_{1,3}}{\mu_{1,1}^2}&&6\frac{\mu_{2,2}}{\mu_{1,1}^2}
&&4\frac{\mu_{3,1}}{\mu_{1,1}^2}&&\frac{\mu_{4,0}}{\mu_{1,1}^2}
\\
&&&&&&\vdots&&&&&&
\\
\frac{\mu_{0, r}}{\mu_{1,1}^{r/2}}&&\left(
\begin{smallmatrix}
r
\\
1
\end{smallmatrix}
\right)\frac{\mu_{1, r-1}}{\mu_{1,1}^{r/2}}&&\cdots&&\left(
\begin{smallmatrix}
r
\\
l
\end{smallmatrix}
\right)\frac{\mu_{l, r-l}}{\mu_{1,1}^{r/2}}&&\cdots&&\left(
\begin{smallmatrix}
r
\\
r-1
\end{smallmatrix}
\right)\frac{\mu_{r-1,1}}{\mu_{1,1}^{r/2}}&&\frac{\mu_{r,0}}{\mu_{1,1}^{r/2}}
\\
\end{array}
\end{gather*}

Observe that the corresponding $n $-th order normalized moment
 $ \hat{m}_n(\theta)=m_n(\theta)/\mu_{1,1}^{\frac{n}{2}}$ is invariant under scaling.
\begin{lemma}[orbit separation property of $T^{2N-2}_{\rm scale}(I)$] Let $I_1$, $I_2 $  be two discrete gray scale images with
the same number $N $  of pixels.
There exists a~scaling $g\in\mathbb{R}_+ $  such that $g\circ I_1=I_2$,  where $\circ $  is defined as
in~\eqref{example2} $\Longleftrightarrow $  $T^r_{\rm scale}(I_1)=T^r_{\rm scale}(I_2) $  for $r\geq2N-2$.
\end{lemma}
\begin{proof} \looseness=-1
 $ \Rightarrow $  If $\exists\,g\in G $  such that $g\circ I_1 = I_2$,  we have $z_k^{(2)} = \lambda
z_k^{(1)} $ and $ \rho_1\big(z_k^{(1)}\big) = \rho_2\big(z_k^{(2)}\big) $ for some $ \lambda \in \mathbb{R}_+$,  $k = 1, \dots, N $.
Since for $I_1 $  and $I_2$,  the corresponding scaling invariant moments are
\begin{gather*}
\hat{\mu}_{j, l}^{(1)}=\frac{\mu_{j, l}^{(1)}}{\big(\mu_{1,1}^{(1)}\big)^{\frac{j+l}{2}}}=\frac{\sum\limits_{k=1}
^N\big(z_k^{(1)}\big)^j\big(\bar{z}_k^{(1)}\big)^l\rho_1\big(z_k^{(1)}\big)}{\left(\sum\limits_{k=1}^Nz_k^{(1)}\bar{z}_k^{(1)}
\rho_1\big(z_k^{(1)}\big)\right)^{\frac{j+l}{2}}},
\\
\hat{\mu}_{j, l}^{(2)}=\frac{\mu_{j, l}^{(2)}}{\big(\mu_{1,1}^{(2)}\big)^{\frac{j+l}{2}}}=\frac{\sum\limits_{k=1}
^N\big(z_k^{(2)}\big)^j\big(\bar{z}_k^{(2)}\big)^l\rho_2\big(z_k^{(2)}\big)}{\left(\sum\limits_{k=1}^Nz_k^{(2)}\bar{z}_k^{(2)}
\rho_2\big(z_k^{(2)}\big)\right)^{\frac{j+l}{2}}}
=\frac{\sum\limits_{k=1}^N\big(\lambda z_k^{(1)}\big)^j\big(\lambda\bar{z}_k^{(1)}\big)^l\rho_1\big(z_k^{(1)}\big)}
{\left(\sum\limits_{k=1} ^N\big(\lambda z_k^{(1)}\big)\big(\lambda\bar{z}_k^{(1)}\big)\rho_1\big(z_k^{(1)}\big)\right)^{\frac{j+l}{2}}}
\\
\hphantom{\hat{\mu}_{j, l}^{(2)}}{}
=\frac{\sum\limits_{k=1}
^N\lambda^{j+l}\big(z_k^{(1)}\big)^j\big(\bar{z}_k^{(1)}\big)^l\rho_1\big(z_k^{(1)}\big)}{\left(\sum\limits_{k=1}^N\lambda^2z_k^{(1)}
\bar{z}_k^{(1)}\rho_1\big(z_k^{(1)}\big)\right)^{\frac{j+l}{2}}}
=\frac{\lambda^{j+l}\sum\limits_{k=1}^N\big(z_k^{(1)}\big)^j\big(\bar{z}_k^{(1)}\big)^l\rho_1\big(z_k^{(1)}\big)}{\lambda^{j+l}\left(
\sum\limits_{k=1}^Nz_k^{(1)}\bar{z}_k^{(1)}\rho_1\big(z_k^{(1)}\big)\right)^{\frac{j+l}{2}}}
\\
\hphantom{\hat{\mu}_{j, l}^{(2)}}{}
=\frac{\sum\limits_{k=1}^N\big(z_k^{(1)}
\big)^j\big(\bar{z}_k^{(1)}\big)^l\rho_1\big(z_k^{(1)}\big)}{\left(\sum\limits_{k=1}^Nz_k^{(1)}\bar{z}_k^{(1)}\rho_1\big(z_k^{(1)}\big)\right)
^{\frac{j+l}{2}}}=\hat{\mu}_{j, l}^{(1)}.
\end{gather*}
Therefore $ T^r_{\rm scale}(I_1) = T^r_{\rm scale}(I_2) $ for any $ r\in\mathbb{Z}_{\geq0}$.

\looseness=-1
 $ \Leftarrow $ If $ T^r_{\rm scale}(I_1) = T^r_{\rm scale}(I_2) $ for $r\geq 2N-2$, from Corollary~\ref{Ttoimage}, we
conclude that $ I_1^{\rm scale} = I_2^{\rm scale}$, i.e.
\begin{gather*}
\big\{\big(z_{k, {\rm scale}}^{(1)}, \rho_1\big(z_{k, {\rm scale}}^{(1)}\big)\big)\big\}_{k=1}^N=\big\{\big(
z_{k, {\rm scale}}^{(2)}, \rho_2\big(z_{k, {\rm scale}}^{(2)}\big)\big)\big\}_{k=1}^N.
\end{gather*}
Hence $ \exists\, \lambda_1, \lambda_2\in \mathbb{R}_+ $ s.t.\
 $ z_{k, {\rm scale}}^{(1)} = \lambda_1z_k^{(1)}$, $z_{k, {\rm scale}}^{(2)} =
\lambda_2z_k^{(2)} $ with $ \rho_1\big(z_{k, {\rm scale}}^{(1)}\big) = \rho_1\big(z_k^{(1)}\big) $ and $ \rho_2\big(z_{k, {\rm scale}}^{(2)}\big) =
\rho_2\big(z_k^{(2)}\big) $ for any $ k = 1,2, \dots, N$.  After relabeling, we have $ z_{k, {\rm scale}}^{(1)} =
z_{k, {\rm scale}}^{(2)} $ and $ \rho_1\big(z_{k, {\rm scale}}^{(1)}\big) = \rho_2\big(z_{k, {\rm scale}}^{(2)}\big)$.  Then $ \exists\,
\frac{\lambda_1}{\lambda_2}\in \mathbb{R}_+ = G $ satisfying
\begin{gather*}
z_k^{(1)}\frac{\lambda_1}{\lambda_2}=z_k^{(2)},
\qquad
\rho_1\left(z_k^{(1)}\frac{\lambda_2}{\lambda_1}\right)=\rho_1\big(z_k^{(1)}\big)=\rho_2\big(z_k^{(2)}\big),
\end{gather*}
i.e.\
 $ \exists\, g=\frac{\lambda_2}{\lambda_1}\in G = \mathbb{R}_+ $ such that $ g\circ I_1 = I_2$.
\end{proof}

More generally, consider the action of group $G $  of diagonal matrices on $\mathbb{R}_+^2 $  by scaling
\begin{gather*}
\left(
\begin{pmatrix}
\lambda_1&0
\\
0&\lambda_2
\end{pmatrix}
,
\begin{pmatrix}
x
\\
y
\end{pmatrix}
\right)\mapsto
\begin{pmatrix}
\lambda_1&0
\\
0&\lambda_2
\end{pmatrix}
\begin{pmatrix}
x
\\
y
\end{pmatrix}
=
\begin{pmatrix}
\lambda_1x
\\
\lambda_2y
\end{pmatrix}
, \\
\forall\,
\begin{pmatrix}
\lambda_1&0
\\
0&\lambda_2
\end{pmatrix}
\in G, \qquad \forall\,
\begin{pmatrix}
x
\\
y
\end{pmatrix}
 \in\mathbb{R}^2.
\end{gather*}
Then the induced transformation on the image $I= \{ (z_k, \rho(z_k) ) \}_{k=1}^N $  is{\samepage
\begin{gather*}
\begin{pmatrix}
\lambda_1&0
\\
0&\lambda_2
\end{pmatrix}
\circ\big\{\big(z_k, \rho(z_k)\big)\big\}_{k=1}^N=\big\{\big(
\lambda_1x_k+i\lambda_2y_k, \rho(z_k)\big)\big\}_{k=1}^N,
\\
\forall\,
\begin{pmatrix}
\lambda_1&0
\\
0&\lambda_2
\end{pmatrix}
\in G, \qquad  z_k=x_k+iy_k.
\end{gather*}
In other words, the image is scaled by a~factor $\lambda_1 $  horizontally and scaled by $\lambda_2 $
vertically.}

Notice that after the transformation, the pixel coordinates become
\begin{gather*}
\hat{z}_k=\lambda_1x_k+i\lambda_2y_k=\lambda_1\frac{z_k+\bar{z}_k}{2}+i\lambda_2\frac{z_k-\bar{z}_k}{2i}
=\frac{\lambda_1+\lambda_2}{2}z_k+\frac{\lambda_1-\lambda_2}{2}\bar{z}_k,
\\
\bar{\hat{z}}_k=\lambda_1x_k-i\lambda_2y_k=\lambda_1\frac{z_k+\bar{z}_k}{2}-i\lambda_2\frac{z_k-\bar{z}_k}
{2i}=\frac{\lambda_1-\lambda_2}{2}z_k+\frac{\lambda_1+\lambda_2}{2}\bar{z}_k.
\end{gather*}

Then we have the transformed complex moments
\begin{gather*}
\hat{\mu}_{j, l}=\sum_{k=1}^N\hat{z}_k^j\bar{\hat{z}}_k^l\rho(\hat{z}_k)=\sum_{k=1}^N\left(
\frac{\lambda_1+\lambda_2}{2}z_k+\frac{\lambda_1-\lambda_2}{2}\bar{z}_k\right)^j\left(
\frac{\lambda_1-\lambda_2}{2}z_k+\frac{\lambda_1+\lambda_2}{2}\bar{z}_k\right)^l\rho(z_k)
\\
\hphantom{\hat{\mu}_{j, l}}{}
=\sum_{k=1}^N\rho(z_k)\left[ \sum_{s=0}^j
\begin{pmatrix}
j
\\
s
\end{pmatrix}
\left(\frac{\lambda_1+\lambda_2}{2}\right)^sz_k^s\left(\frac{\lambda_1-\lambda_2}{2}\right)^{j-s}\bar{z}_k^{j-s}\right]
\\
\hphantom{\hat{\mu}_{j, l}=}{} \times
\left[ \sum_{t=0}^l
\begin{pmatrix}
l
\\
t
\end{pmatrix}
\left(\frac{\lambda_1-\lambda_2}{2}\right)^tz_k^t\left(\frac{\lambda_1+\lambda_2}{2}\right)^{l-t}\bar{z}_k^{l-t}\right]
\\
\hphantom{\hat{\mu}_{j, l} }{}
=\sum_{k=1}^N\rho(z_k)\left[ \sum_{s=0}^j\sum_{t=0}^l
\begin{pmatrix}
j
\\
s
\end{pmatrix}
\begin{pmatrix}
l
\\
t
\end{pmatrix}
\left(\frac{\lambda_1+\lambda_2}{2}\right)^{l-t+s}\left(\frac{\lambda_1-\lambda_2}{2}\right)^{j-s+t}z_k^{s+t}\bar{z}
_k^{j+l-s-t}\right]
\\
\hphantom{\hat{\mu}_{j, l} }{}
=\sum_{s=0}^j\sum_{t=0}^l
\begin{pmatrix}
j
\\
s
\end{pmatrix}
\begin{pmatrix}
l
\\
t
\end{pmatrix}
\left(\frac{\lambda_1+\lambda_2}{2}\right)^{l-t+s}\left(\frac{\lambda_1-\lambda_2}{2}\right)^{j-s+t}\left[ \sum_{k=1}
^N\rho(z_k)z_k^{s+t}\bar{z}_k^{j+l-s-t}\right]
\\
\hphantom{\hat{\mu}_{j, l} }{}
=\frac{1}{2^{j+l}}\sum_{s=0}^j\sum_{t=0}^l
\begin{pmatrix}
j
\\
s
\end{pmatrix}
\begin{pmatrix}
l
\\
t
\end{pmatrix}
(\lambda_1+\lambda_2)^{l-t+s}(\lambda_1-\lambda_2)^{j-s+t}\mu_{s+t, j+l-s-t},
\qquad
\forall\,  j, l\in\mathbb{Z}_{\geq0},
\end{gather*}
which is a~linear combination of the last row of the Pascal triangle $T^{j+l}(I) $.
\end{example}
\begin{example}
\label{exp3}
Consider the action of $G=\{z\in\mathbb{C}||z|=1\}$ on $\mathbb{C}$ by rotation
\begin{gather*}
\big(e^{i\theta_0}, z\big)\mapsto ze^{i\theta_0}, \qquad \forall\,  e^{i\theta_0}\in G, \qquad \forall\,  z\in\mathbb{C}.
\end{gather*}
Then the induced transformation on the image $I= \{ (z_k, \rho(z_k) ) \}_{k=1}^N $  is
\begin{gather}
\label{example3}
e^{i\theta_0}\circ \{ (z_k, \rho(z_k) ) \}_{k=1}^N= \big\{ \big(e^{i\theta_0}
z_k, \rho(z_k)\big)\big\}_{k=1}^N,
\qquad
\forall\,  e^{i\theta_0}\in G.
\end{gather}
In other words, the image is rotated counterclockwise with an angle $\theta_0 $.
The transformed complex moments are
\begin{gather*}
\mu'_{j, l}=\sum_{k=1}^N{z'}_k^j\bar{z'}_k^l\rho(z'_k)=\sum_{k=1}^N\big(z_ke^{i\theta_0}\big)^j\big(\bar{z}
_ke^{-i\theta_0}\big)^l\rho(z_k)
=\sum_{k=1}^N e^{i(j-l)\theta_0}z_k^j\bar{z}_k^l\rho(z_k)\\
\hphantom{\mu'_{j, l}}{}
=e^{i(j-l)\theta_0}\sum_{k=1}^Nz_k^j\bar{z}
_k^l\rho(z_k)
=\mu_{j, l}e^{i(j-l)\theta_0},
\qquad
\forall\,  j, l\in\mathbb{Z}_{\geq0}.
\end{gather*}
Written in matrix form, the moment matrix for the new image $I' $  after rotation is
\begin{gather*}
\tau_N(I')=
\begin{pmatrix}
1&0&0&\cdots&0
\\
0&e^{-i\theta_0}&0&\cdots&0
\\
0&0&e^{-i2\theta_0}&\cdots&0
\\
\vdots&\vdots&\vdots&\ddots&\vdots
\\
0&\cdots&\cdots&0&e^{-i(N-1)\theta_0}
\end{pmatrix}
\tau_N(I)
\begin{pmatrix}
1&0&0&\cdots&0
\\
0&e^{i\theta_0}&0&\cdots&0
\\
0&0&e^{i2\theta_0}&\cdots&0
\\
\vdots&\vdots&\vdots&\ddots&\vdots
\\
0&\cdots&\cdots&0&e^{i(N-1)\theta_0}
\end{pmatrix}
.
\end{gather*}
Now we will use the moving frame method of Fels and Olver to obtain a~set of invariant functions of the
moments.
If $\mu_{0,2}\neq0$,  we normalize the imaginary part of $\mu'_{0,2}$ to zero by specifying the rotation
angle $\theta_0$.  Since $\mu'_{0,2}=\mu_{0,2}e^{-i2\theta_0}$,  looking at $\vec{\mu}_{0,2}$ as a~vector in
 $ \mathbb{R}^2 $  representing the complex number $\mu_{0,2}$,  we set
\begin{gather*}
2\theta_0=\sphericalangle(\vec{\mu}_{0,2}, \vec{e}_1)+2k\pi, \qquad k\in\mathbb{Z}.
\end{gather*}
Here $\sphericalangle(\vec{x}, \vec{y})=\tan^{-1}\big(\frac{y_2}{y_1}\big)-\tan^{-1}\big(\frac{x_2}{x_1}\big)\in(-\pi, \pi] $
denotes the angle from $\vec{x}$ to $\vec{y}$,  $\vec{e}_1=(1,0)^T $  is one of the standard basis of
 $ \mathbb{R}^2$.  The real part of $\mu'_{0,2}$ then reduces to its magnitude $|\mu_{0,2}|$.

Since $\theta_0\in(-\pi, \pi]$, $2\theta_0\in(-2\pi,2\pi]$.  To uniquely determine the value of $\theta_0$,  we
consider $\mu_{1,2}'=\mu_{1,2}e^{-i\theta_0}$.  We choose $\theta_0 $  such that $\text{Re}(\mu'_{1,2})\geq0 $,
which leads to the moving frame formulae
\begin{gather}
\label{theta0}
\theta_0=
\begin{cases}
\frac{1}{2}\sphericalangle(\vec{\mu}_{0,2}, \vec{e}_1)
&\text{if}\ \
\sphericalangle(\vec{\mu}_{1,2}, \vec{e}_1)-\frac{1}{2}\sphericalangle(\vec{\mu}_{0,2}, \vec{e}_1)\in \big[{-}\frac{\pi}{2}, \frac{\pi}{2}\big],
\\
\frac{1}{2}\sphericalangle(\vec{\mu}_{0,2}, \vec{e}_1)+\pi
&\text{if}\ \
\sphericalangle(\vec{\mu}_{1,2}, \vec{e}_1)-\frac{1}{2}\sphericalangle(\vec{\mu}_{0,2}, \vec{e}_1)\in\big(\frac{\pi}{2}, \frac{3\pi}{2}\big],
\\
\frac{1}{2}\sphericalangle(\vec{\mu}_{0,2}, \vec{e}_1)-\pi
&\text{if}\ \
\sphericalangle(\vec{\mu}_{1,2}, \vec{e}_1)-\frac{1}{2}\sphericalangle(\vec{\mu}_{0,2}, \vec{e}_1)\in\big[{-}\frac{3\pi}{2},-\frac{\pi}{2}\big).
\end{cases}
\end{gather}

By applying the moving frame to the moment matrix, we obtain the matrix
\begin{gather*}
\tau'_N(I)=
\begin{pmatrix}
1&0&0&\cdots&0
\\
0&e^{-i\theta_0}&0&\cdots&0
\\
0&0&e^{-i2\theta_0}&\cdots&0
\\
\vdots&\vdots&\vdots&\ddots&\vdots
\\
0&\cdots&\cdots&0&e^{-i(N-1)\theta_0}
\end{pmatrix}
\tau_N(I)
\begin{pmatrix}
1&0&0&\cdots&0
\\
0&e^{i\theta_0}&0&\cdots&0
\\
0&0&e^{i2\theta_0}&\cdots&0
\\
\vdots&\vdots&\vdots&\ddots&\vdots
\\
0&\cdots&\cdots&0&e^{i(N-1)\theta_0}
\end{pmatrix}
,
\end{gather*}
with $\theta_0 $  satisfying~\eqref{theta0}.
By equivariance of the moving frame, all the entries of $\tau'_N(I) $  are invariant under rotation.

By normalizing the coordinates of $T^r(I)$,  we obtain the rotational invariant Pascal triangle
 $ T^r_{\rm rotate}(I) $  for a~discrete image $I $:
\begin{gather*}
\begin{array}{@{}c@{\,}c@{\,}c@{\,}c@{\,}c@{\,}c@{\,}c@{\,}c@{\,}c@{\,}c@{\,}c@{\,}c@{\,}c@{}}&&&&&&\mu_{0,0}&&&&&&
\\
&&&&&\frac{\mu_{0,1}}{e^{i\theta_0}}&&\frac{\mu_{1,0}}{e^{-i\theta_0}}&&&&&
\\
&&&&|\mu_{0,2}|&&2\mu_{1,1}&&|\mu_{2,0}|&&&&
\\
&&&\frac{\mu_{0,3}}{e^{i3\theta_0}}&&3\frac{\mu_{1,2}}{e^{i\theta_0}}&&3\frac{\mu_{2,1}}{e^{-i\theta_0}}
&&\frac{\mu_{3,0}}{e^{-i3\theta_0}}&&&
\\
&&\frac{\mu_{0,4}}{e^{i4\theta_0}}&&4\frac{\mu_{1,3}}{e^{i2\theta_0}}&&\mu_{2,2}&&4\frac{\mu_{3,1}}
{e^{-i2\theta_0}}&&\frac{\mu_{4,0}}{e^{-i4\theta_0}}
\\
&&&&&&\vdots&&&&&&
\\
&\frac{\mu_{0, r}}{e^{ir\theta_0}}&&&\cdots&&\left(
\begin{smallmatrix}
r
\\
l
\end{smallmatrix}
\right)\frac{\mu_{l, r-l}}{e^{i(r-2l)\theta_0}}&&\cdots&&&\frac{\mu_{r,0}}{e^{-ir\theta_0}}&
\end{array}
\end{gather*}

Observe that the corresponding $n $-th order moment $m'_n(\theta)=m_n(\theta-\theta_0)$,  with $\theta_0 $
def\/ined as in~\eqref{theta0}, is invariant under rotations.
\begin{lemma}[orbit separation property of $T^{2N-2}_{\rm rotate}(I) $ ] Let $I_1$, $I_2 $  be two discrete gray scale images with
the same number $N $  of pixels.
There exists a~rotation $g\in\{z\in\mathbb{C}||z|=1\}$ such that $g\circ I_1=I_2$,  where $\circ $  is
defined as in~\eqref{example3} $\Longleftrightarrow $  $T^r_{\rm rotate}(I_1)=T^r_{\rm rotate}(I_2) $  for
 $ r\geq2N-2 $.
\end{lemma}

\begin{proof}
 $ \Rightarrow $ If $ \exists\,g\in G $ such that $ g\circ I_1 = I_2$, we have $ z_k^{(2)} =
z_k^{(1)}e^{i\theta_0} $ and $ \rho_1(z_k^{(1)}) = \rho_2(z_k^{(2)})$, for some $ e^{i\theta_0} \in
\{z\in\mathbb{C}\,|\, |z| = 1\}$, $k = 1, \dots, N$.
For $I_1 $ and $ I_2$, the corresponding scaling invariant moments are
\begin{gather*}
\hat{\mu}_{j, l}^{(1)}=\mu_{j, l}^{(1)}e^{-i(l-j)\theta_1}=\sum_{k=1}^N\big(z_k^{(1)}\big)^j\big(\bar{z}_k^{(1)}
\big)^l\rho_1\big(z_k^{(1)}\big)e^{i(j-l)\theta_1},
\\
\hat{\mu}_{j, l}^{(2)}=\mu_{j, l}^{(2)}e^{-i(l-j)\theta_2}=\sum_{k=1}^N\big(z_k^{(2)}\big)^j\big(\bar{z}_k^{(2)}
\big)^l\rho_2\big(z_k^{(2)}\big)e^{i(j-l)\theta_2}
\\
\hphantom{\hat{\mu}_{j, l}^{(2)}}{}
=\sum_{k=1}^N\big(z_k^{(1)}e^{i\theta_0}\big)^j\big(\bar{z}_k^{(1)}e^{-i\theta_0}\big)^l\rho_1\big(z_k^{(1)}\big)e^{i(j-l)\theta_2}
=\sum_{k=1}^N\big(z_k^{(1)}\big)^j\big(\bar{z}_k^{(1)}\big)^l\rho_1\big(z_k^{(1)}\big)e^{i(j-l)(\theta_0+\theta_2)}.
\end{gather*}
Since $ \mu_{0,2}^{(2)} = \mu_{0,2}^{(1)}e^{-i2\theta_0}$,
$\mu_{1,2}^{(2)} = \mu_{1,2}^{(1)}e^{-i\theta_0}$, we have
\begin{gather}
\sphericalangle\big(\vec{\mu}_{0,2}^{(2)}, \vec{e}_1\big)=\sphericalangle\big(\vec{\mu}_{0,2}^{(1)}, \vec{e}
_1\big)-2\theta_0+2k_1\pi,
\nonumber\\
\sphericalangle\big(\vec{\mu}_{1,2}^{(2)}, \vec{e}_1\big)=\sphericalangle\big(\vec{\mu}_{1,2}^{(1)}, \vec{e}
_1\big)-\theta_0+2k_2\pi,
\qquad
k_1, k_2\in\mathbb{Z}.\label{angle}
\end{gather}
Notice that $ \sphericalangle\big(\vec{\mu}_{0,2}^{(l)}, \vec{e}_1\big)$,
$\sphericalangle\big(\vec{\mu}_{1,2}^{(l)}, \vec{e}_1\big)$, $\theta_0 \in (-\pi, \pi]$, for $l = 1,2$.  Hence $ k_1, k_2 =
0, \pm 1$.  To decide~$ \theta_1 $ and~$ \theta_2$, we consider
\begin{gather}
\label{relation12}
\sphericalangle\big(\vec{\mu}_{1,2}^{(2)}, \vec{e}_1\big)-\frac{1}{2}\sphericalangle\big(\vec{\mu}_{0,2}^{(2)}, \vec{e}
_1\big)=\sphericalangle\big(\vec{\mu}_{1,2}^{(1)}, \vec{e}_1\big)-\frac{1}{2}\sphericalangle\big(\vec{\mu}_{0,2}^{(1)}
, \vec{e}_1\big)+(2k_2-k_1)\pi
\end{gather}
by using~\eqref{angle}.

If $ \sphericalangle\big(\vec{\mu}_{1,2}^{(2)}, \vec{e}_1\big) -
\frac{1}{2}\sphericalangle\big(\vec{\mu}_{0,2}^{(2)}, \vec{e}_1\big)\in
\big[{-}\frac{\pi}{2}, \frac{\pi}{2}\big]$, from~\eqref{theta0} we know that $ \theta_2 =
\frac{1}{2}\sphericalangle\big(\vec{\mu}_{0,2}^{(2)}, \vec{e}_1\big) $.
Since $ \sphericalangle\big(\vec{\mu}_{1,2}^{(1)}, \vec{e}_1\big) -
\frac{1}{2}\sphericalangle\big(\vec{\mu}_{0,2}^{(1)}, \vec{e}_1\big) \in
\big[{-}\frac{3\pi}{2}, \frac{3\pi}{2}\big]$, then $ 2k_2-k_1 $ is either $ 0 $ or $ \pm1 $ in~\eqref{relation12}.
If $ k_1 = 0, k_2 = 0$, there is $ \sphericalangle\big(\vec{\mu}_{1,2}^{(1)}, \vec{e}_1\big) -
\frac{1}{2}\sphericalangle\big(\vec{\mu}_{0,2}^{(1)}, \vec{e}_1\big) \in \big[{-}\frac{\pi}{2}, \frac{\pi}{2}\big] $.
Hence $ \theta_1 = \frac{1}{2}\sphericalangle\big(\vec{\mu}_{0,2}^{(1)}, \vec{e}_1\big)$.  Then from~\eqref{angle} we
have $ \theta_2 = \theta_1 - \theta_0$.  Thus $ \hat{\mu}_{j, l}^{(2)} =
\sum\limits_{k=1}^N\big(z_k^{(1)}\big)^j\big(\bar{z}_k^{(1)}\big)^l\rho_1\big(z_k^{(1)}\big)e^{i(j-l)\theta_1} = \hat{\mu}_{j, l}^{(1)} $.
If $ k_1 = 1$, $k_2 = 0$, there is $ \sphericalangle\big(\vec{\mu}_{1,2}^{(1)}, \vec{e}_1\big) -
\frac{1}{2}\sphericalangle\big(\vec{\mu}_{0,2}^{(1)}, \vec{e}_1\big) \in \big(\frac{\pi}{2}, \frac{3\pi}{2}\big] $.
Hence $ \theta_1 = \frac{1}{2}\sphericalangle\big(\vec{\mu}_{0,2}^{(1)}, \vec{e}_1\big)+\pi$.  Then
from~\eqref{angle} we still have $ \theta_2 = \theta_1 - \theta_0$.  Thus $ \hat{\mu}_{j, l}^{(2)} =
\hat{\mu}_{j, l}^{(1)}$.
If $ k_1 = 1$, $k_2 = 1$, there is $ \sphericalangle\big(\vec{\mu}_{1,2}^{(1)}, \vec{e}_1\big) -
\frac{1}{2}\sphericalangle\big(\vec{\mu}_{0,2}^{(1)}, \vec{e}_1\big) \in \big[{-}\frac{3\pi}{2},-\frac{\pi}{2}\big) $.
Hence $ \theta_1 = \frac{1}{2}\sphericalangle\big(\vec{\mu}_{0,2}^{(1)}, \vec{e}_1\big)-\pi$.  Then
from~\eqref{angle} we have $ \theta_2 = \theta_1 - \theta_0+2\pi$.  Thus $ \hat{\mu}_{j, l}^{(2)} =
\sum\limits_{k=1}^N\big(z_k^{(1)}\big)^j\big(\bar{z}_k^{(1)}\big)^l\rho_1\big(z_k^{(1)}\big)e^{i(j-l)\theta_1} = \hat{\mu}_{j, l}^{(1)} $.

If $ \sphericalangle\big(\vec{\mu}_{1,2}^{(2)}, \vec{e}_1\big) -
\frac{1}{2}\sphericalangle\big(\vec{\mu}_{0,2}^{(2)}, \vec{e}_1\big)\in
\big(\frac{\pi}{2}, \frac{3\pi}{2}\big]$, from~\eqref{theta0} we know that $ \theta_2 =
\frac{1}{2}\sphericalangle\big(\vec{\mu}_{0,2}^{(2)}, \vec{e}_1\big)+\pi $.
Since $ \sphericalangle\big(\vec{\mu}_{1,2}^{(1)}, \vec{e}_1\big) -
\frac{1}{2}\sphericalangle\big(\vec{\mu}_{0,2}^{(1)}, \vec{e}_1\big) \in
\big[{-}\frac{3\pi}{2}, \frac{3\pi}{2}\big]$, then $ 2k_2-k_1 = 0,1,2 $ in~\eqref{relation12}.
If $ k_1 = 1, k_2 = 1$, there is $ \sphericalangle\big(\vec{\mu}_{1,2}^{(1)}, \vec{e}_1\big) -
\frac{1}{2}\sphericalangle\big(\vec{\mu}_{0,2}^{(1)}, \vec{e}_1\big) \in \big[{-}\frac{\pi}{2}, \frac{\pi}{2}\big] $.
Hence $ \theta_1 = \frac{1}{2}\sphericalangle\big(\vec{\mu}_{0,2}^{(1)}, \vec{e}_1\big)$.  Then from~\eqref{angle} we
have $ \theta_2 -\pi = \theta_1 - \theta_0 + \pi$.  Thus $ \hat{\mu}_{j, l}^{(2)} =
\sum\limits_{k=1}^N\big(z_k^{(1)}\big)^j\big(\bar{z}_k^{(1)}\big)^l\rho_1\big(z_k^{(1)}\big)e^{i(j-l)\theta_1} = \hat{\mu}_{j, l}^{(1)} $.
If $ k_1 = 0$, $k_2 = 0$, there is $ \sphericalangle\big(\vec{\mu}_{1,2}^{(1)}, \vec{e}_1\big) -
\frac{1}{2}\sphericalangle\big(\vec{\mu}_{0,2}^{(1)}, \vec{e}_1\big) \in \big(\frac{\pi}{2}, \frac{3\pi}{2}\big] $.
Hence $ \theta_1 = \frac{1}{2}\sphericalangle\big(\vec{\mu}_{0,2}^{(1)}, \vec{e}_1\big)+\pi$.  Then from~\eqref{angle} we
have $ \theta_2 - \pi = \theta_1 - \theta_0 - \pi$.  Thus $ \hat{\mu}_{j, l}^{(2)} = \hat{\mu}_{j, l}^{(1)} $.
If $ k_1 = 0$, $k_2 = 1$, there is $ \sphericalangle\big(\vec{\mu}_{1,2}^{(1)}, \vec{e}_1\big) -
\frac{1}{2}\sphericalangle\big(\vec{\mu}_{0,2}^{(1)}, \vec{e}_1\big) \in \big[{-}\frac{3\pi}{2},-\frac{\pi}{2}\big) $.
Hence $ \theta_1 = \frac{1}{2}\sphericalangle\big(\vec{\mu}_{0,2}^{(1)}, \vec{e}_1\big)-\pi$.  Then
from~\eqref{angle} we have $ \theta_2 - \pi = \theta_1 - \theta_0+\pi$.  Thus $ \hat{\mu}_{j, l}^{(2)} =
\sum\limits_{k=1}^N\big(z_k^{(1)}\big)^j\big(\bar{z}_k^{(1)}\big)^l\rho_1\big(z_k^{(1)}\big)e^{i(j-l)\theta_1} = \hat{\mu}_{j, l}^{(1)} $.

If $ \sphericalangle\big(\vec{\mu}_{1,2}^{(2)}, \vec{e}_1\big) -
\frac{1}{2}\sphericalangle\big(\vec{\mu}_{0,2}^{(2)}, \vec{e}_1\big)\in \big[{-}\frac{3\pi}{2},-\frac{\pi}{2}\big)$, through
the similar discussion, we can still conclude that $ \hat{\mu}_{j, l}^{(2)} = \hat{\mu}_{j, l}^{(1)} $.
Therefore $ T^r_{\rm scale}(I_1) = T^r_{\rm scale}(I_2) $ for any $ r\in\mathbb{Z}_{\geq 0}$.

 $ \Leftarrow $ If $ T^r_{\rm rotate}(I_1) = T^r_{\rm rotate}(I_2) $ for $r\geq 2N-2$, from Corollary~\ref{Ttoimage}, we
conclude that $ I_1^{\rm rotate} = I_2^{\rm rotate}$, i.e.
\begin{gather*}
\big\{\big(z_{k, {\rm rotate}}^{(1)}, \rho_1\big(z_{k, {\rm rotate}}^{(1)}\big)\big)\big\}_{k=1}^N=\big\{\big(
z_{k, {\rm rotate}}^{(2)}, \rho_2\big(z_{k, {\rm rotate}}^{(2)}\big)\big)\big\}_{k=1}^N.
\end{gather*}
Hence $ \exists\, e^{i\theta_1}, e^{i\theta_2} \in \{z\in\mathbb{C}\,|\, |z| = 1\} $ s.t.\
 $ z_{k, {\rm rotate}}^{(1)} = z_k^{(1)}e^{i\theta_1}$, $z_{k, {\rm rotate}}^{(2)} =
z_k^{(2)}e^{i\theta_2} $ with $ \rho_1\big(z_{k, {\rm rotate}}^{(1)}\big) = \rho_1\big(z_k^{(1)}\big) $ and $ \rho_2\big(z_{k, {\rm rotate}}^{(2)}\big) =
\rho_2\big(z_k^{(2)}\big) $ for any $ k = 1,2, \dots, N$.  Without loss of generality, we assume $ z_{k, {\rm rotate}}^{(1)} =
z_{k, {\rm rotate}}^{(2)} $ and $ \rho_1\big(z_{k, {\rm rotate}}^{(1)}\big) = \rho_2\big(z_{k, {\rm rotate}}^{(2)}\big)$.  Then $ \exists\,
e^{i(\theta_1-\theta_2)}\in \{z\in\mathbb{C}\,|\,|z| = 1\} = G $ satisfying
\begin{gather*}
z_k^{(1)}e^{i(\theta_1-\theta_2)}=z_k^{(2)},
\qquad
\rho_1\big(z_k^{(1)}e^{i(\theta_1-\theta_2)}\big)=\rho_1\big(z_{k}^{(1)}\big)=\rho_2\big(z_k^{(2)}\big),
\end{gather*}
i.e.\
 $ \exists\, g=e^{i(\theta_1-\theta_2)}\in G = \{z\in\mathbb{C}\, |\, |z| = 1\} $ such that $ g\circ I_1 = I_2$.
\end{proof}
\end{example}

In conclusion, we have the translation, scaling and rotation invariant Pascal triangle $T^r_{\rm int}(I) $:
\begin{gather*}
\begin{array}{@{}c@{\,}c@{\,}c@{\,}c@{\,}c@{\,}c@{\,}c@{\,}c@{\,}c@{\,}c@{\,}c@{}}&&&&&\tilde{\mu}_{0,0}&&&&&
\\
&&&&\frac{\tilde{\mu}_{0,1}e^{-i\theta_0}}{\sqrt{\tilde{\mu}_{1,1}}}&&\frac{\tilde{\mu}_{1,0}e^{i\theta_0}}
{\sqrt{\tilde{\mu}_{1,1}}}&&&&
\\
&&&\frac{|\tilde{\mu}_{0,2}|}{\tilde{\mu}_{1,1}}&&2&&\frac{|\tilde{\mu}_{2,0}|}{\tilde{\mu}_{1,1}}&&&
\\
&&\frac{\tilde{\mu}_{0,3}e^{-i3\theta_0}}{\tilde{\mu}_{1,1}^{3/2}}&&3\frac{\tilde{\mu}_{1,2}e^{-i\theta_0}}
{\tilde{\mu}_{1,1}^{3/2}}&&3\frac{\tilde{\mu}_{2,1}e^{i\theta_0}}{\tilde{\mu}_{1,1}^{3/2}}
&&\frac{\tilde{\mu}_{3,0}e^{i3\theta_0}}{\tilde{\mu}_{1,1}^{3/2}}&&
\\
&&&&&\vdots&&&&&
\\
\frac{\tilde{\mu}_{0, r}e^{-ir\theta_0}}{\tilde{\mu}_{1,1}^{r/2}}&&&\cdots&&\frac{\left(
\begin{smallmatrix}
r
\\
l
\end{smallmatrix}
\right)\tilde{\mu}_{l, r-l}}{\tilde{\mu}_{1,1}^{r/2}e^{i(r-2l)\theta_0}}&&\cdots&&&\frac{\tilde{\mu}_{r,0}
e^{ir\theta_0}}{\tilde{\mu}_{1,1}^{r/2}}
\end{array}
\end{gather*}

\section{Geometric interpretation of the moments} \subsection{Shape elongation}

Intuitively, the elongation
of a~shape is described by the relationship between its length and its width.
It is thus a~property that is invariant under translation, scaling and rotation of the image.
To characterize the shape elongation, we therefore consider the invariantization
 $ \hat{\tilde{m}}'_2(\theta) $  of the second order moment $m_2(\theta) $  with respect to translation, scaling
and rotation.
By combining the results of Examples~\ref{exp1},~\ref{exp2} and~\ref{exp3}, we have
\begin{gather}
\hat{\tilde{m}}'_2(\theta)=\frac{1}{4}\left(\frac{|\tilde{\mu}_{0,2}|}{\tilde{\mu}_{1,1}}e^{i2\theta}
+2+\frac{|\tilde{\mu}_{2,0}|}{\tilde{\mu}_{1,1}}e^{-i2\theta}\right) \nonumber
\\
\hphantom{\hat{\tilde{m}}'_2(\theta)}{}
=\frac{1}{4}\left(\frac{|\tilde{\mu}_{0,2}|}{\tilde{\mu}_{1,1}}\big( \cos(2\theta)+i\sin(2\theta)\big)
+2+\frac{|\tilde{\mu}_{2,0}|}{\tilde{\mu}_{1,1}}\big(\cos(2\theta)-i\sin(2\theta)\big)\right)\nonumber
\\
\hphantom{\hat{\tilde{m}}'_2(\theta)}{}
=\frac{1}{4}\left(2\frac{|\tilde{\mu}_{0,2}|}{\tilde{\mu}_{1,1}}\cos(2\theta)+2\right)
=\frac{1}{2}\left(\frac{|\tilde{\mu}_{0,2}|}{\tilde{\mu}_{1,1}}\cos(2\theta)+1\right),\label{m2inv}
\end{gather}
where
\begin{gather*}
\tilde{\mu}_{1,1}=\sum_{k=1}^N(z_k-z_0)(\bar{z}_k-\bar{z}_0)\rho(z_k)=\sum_{k=1}^N\big(
(x_k-x_0)^2+(y_k-y_0)^2\big)\rho(z_k)>0.
\end{gather*}

Recall that when making the moments scaling invariant, we normalized $\hat{\tilde{\mu}}_{1,1}$ to $1 $  by
dividing each $\tilde{\mu}_{j, l}$ by the corresponding power of $\sqrt{\tilde{\mu}_{1,1}} $.
Hence $\sqrt{\tilde{\mu}_{1,1}}$ represents the scale of the image.

Equation~\eqref{m2inv} indicates that the quantity $\frac{|\tilde{\mu}_{0,2}|}{\tilde{\mu}_{1,1}} $
prescribes the relationship between the maximum and the minimum values of the standard deviation of the
random transform.
It thus gives us a~quantif\/ication of the elongation of the shape illustrated by the image\footnote{A.R.~Rostampour et al.~\cite{Proj} previously introduced the $\frac{n_{04}}{n_{02}^2}$ as a~measure of elongation of the
projections of an image.
We observe that our measure of elongation has a~lower order, and that it can be obtained without
normalizing the angle of the image.}.
\begin{lemma}
\label{m01}
If not all $\rho(z_k) $  are zero, then $0\leq\frac{|\tilde{\mu}_{0,2}|}{\tilde{\mu}_{1,1}}\leq1$.
\end{lemma}
\begin{proof}
By def\/inition, $|\tilde{\mu}_{0,2}|\geq0 $  and $\tilde{\mu}_{1,1}>0$,  hence
 $ \frac{|\tilde{\mu}_{0,2}|}{\tilde{\mu}_{1,1}}\geq0$.
If $\frac{|\tilde{\mu}_{0,2}|}{\tilde{\mu}_{1,1}}>1$,  since
 $ \hat{\tilde{m}}'_2(\theta)=\frac{1}{2}\big(\frac{|\tilde{\mu}_{0,2}|}{\tilde{\mu}_{1,1}}\cos(2\theta)+1\big)$,  and
 $ \hat{\tilde{m}}'_2(\theta)\geq0 $  by def\/inition, if we choose $\theta=\frac{\pi}{2}$,  then
 $ \frac{|\tilde{\mu}_{0,2}|}{\tilde{\mu}_{1,1}}\cos(2\theta)+1<0$,  which is a~contradiction.
Therefore $\frac{|\tilde{\mu}_{0,2}|}{\tilde{\mu}_{1,1}}\leq1$.
\end{proof}

The case $\frac{|\tilde{\mu}_{0,2}|}{\tilde{\mu}_{1,1}}=1 $  corresponds to the most extreme elongation,
namely the straight lines.
\begin{lemma}
The pixel coordinates $z_k$  lie on a~single straight line if and only if
 $ \frac{|\tilde{\mu}_{0,2}|}{\tilde{\mu}_{1,1}}=1$.
\end{lemma}
\begin{proof}
 $ \Rightarrow $ Suppose we have a~straight line.
As the line is put vertically, the projection of the line is a~dot.
Hence the second moment of the Radon transform is zero at that angle $ \theta^\ast$, i.e.
\begin{gather}
\label{m2=0}
\hat{\tilde{m}}'_2(\theta^\ast)=\frac{1}{2}\left(\frac{|\tilde{\mu}_{0,2}|}{\tilde{\mu}_{1,1}}
\cos(2\theta^\ast)+1\right)=0.
\end{gather}
Since $-1\leq \cos(2\theta^\ast)\leq 1 $ and from Lemma~\ref{m01} we know
that $ 0\leq\frac{|\tilde{\mu}_{0,2}|}{\tilde{\mu}_{1,1}} \leq 1 $ for any image, we can conclude
that~\eqref{m2=0} is true only when $ \frac{|\tilde{\mu}_{0,2}|}{\tilde{\mu}_{1,1}} = 1 $ and $ \cos(2\theta^\ast)=-1$.
Hence $ \frac{|\tilde{\mu}_{0,2}|}{\tilde{\mu}_{1,1}} = 1 $ is true.

 $ \Leftarrow $ Now suppose $ \frac{|\tilde{\mu}_{0,2}|}{\tilde{\mu}_{1,1}} = 1$.  Combine the results in
Examples~\ref{exp1},~\ref{exp2} and~\ref{exp3}, we conclude that
\begin{gather*}
\hat{\tilde{m}}'_2(\theta)=\tilde{m}_2(\theta-\theta_0)/\tilde{\mu}_{1,1}=\frac{1}{2}\left(\frac{|\tilde{\mu}
_{0,2}|}{\tilde{\mu}_{1,1}}\cos(2\theta)+1\right),
\qquad
\forall\, \theta\in\left(-\frac{\pi}{2}, \frac{\pi}{2}\right],
\end{gather*}
where $ \theta_0 $ satisf\/ies~\eqref{theta0}.
Since $ \hat{\tilde{m}}'_n(\frac{\pi}{2}) = 0$, then there is $ \tilde{\theta} = \frac{\pi}{2} - \theta_0 $ such
that at this angle the centralized second order moment $ \tilde{m}_2(\tilde{\theta}) $ of the image is zero.

Let the Radon transform $ f_{\tilde{\theta}}(r) $ at angle $ \tilde{\theta} $ be a~discrete function.
Without loss of generality, we assume that $ \sum\limits_{k = 1}^{N}f_{\tilde{\theta}} (r_k(\theta) ) =
1 $ and $ f_{\tilde{\theta}} (r_k(\theta) )\geq 0$.
Since
\begin{gather*}
\tilde{m}_2(\tilde{\theta})=\sum_{k=1}^{N}\big(r_k(\tilde{\theta})-r_0(\tilde{\theta})\big)^2f_{\tilde{\theta}}
 (r_k(\theta) )=0,
\end{gather*}
where $ r_0(\tilde{\theta}) $ is the projection of the centroid of the image, we observe
that $ f_{\tilde{\theta}}(r) = \delta(r-r_0(\tilde{\theta}))$.
Then we conclude that the image lies on a~line through $ r_0(\tilde{\theta}) $ with
angle $ \tilde{\theta}+\frac{\pi}{2} $ to the $x$-axis.
\end{proof}

The other extreme case is when $ \frac{|\tilde{\mu}_{0,2}|}{\tilde{\mu}_{1,1}} = 0 $.
This corresponds to $ \hat{\tilde{m}}'_2(\theta) = \text{const}$. So the standard deviation of the
projection is the same for all directions.
There are many ways for this to happen.
One interesting case is the discrete analogue of rotation symmetries.
\begin{definition}
An object is said to have $ \tilde{N} $-fold rotation symmetry ($\tilde{N}$-FRS) if it is unchanged by
a~rotation around its centroid by $ \frac{2k\pi}{\tilde{N}}$, for all $ k = 1, \dots, \tilde{N}$.
\end{definition}

\begin{lemma}
If the image $ I $ has an $ \tilde{N} $-FRS with $ \tilde{N}>2$, then $ \frac{|\tilde{\mu}_{0,2}|}{\tilde{\mu}_{1,1}}=0$.
\end{lemma}

\begin{proof}
Suppose the data have $ \tilde{N} $-FRS.
For a~certain point with distance $ r_k$, $k=1,2, \dots, M$, from the centroid, angle $ \theta_k $ with $x$-axis and
weight $ \rho(z_k)$, there are $ \tilde{N}_k-1 $ more points with the same distance from the centroid and the same
weight $ \rho(z_k) $ but having angle $ \theta_k+ \frac{2j\pi}{\tilde{N}_k}$,
$j = 1, \dots, \tilde{N}_k-1 $ with $x$-axis respectively.
Then the moment $ \tilde{\mu}_{0,2} $ can be written as
\begin{gather*}
\tilde{\mu}_{0,2}=\sum_{k=1}^M\rho(z_k)\left(\sum_{j=0}^{\tilde{N}_k-1}\left(
r_k\cos\left(\theta_k+\frac{2j\pi}{\tilde{N}_k}\right)-ir_k\sin\left(\theta_k+\frac{2j\pi}{\tilde{N}_k}\right)\right)
^2\right)
\\
\hphantom{\tilde{\mu}_{0,2}}{}
=\sum_{k=1}^M\rho(z_k)\left(\sum_{j=0}^{\tilde{N}_k-1}r_k^2\cos\left(2\theta_k+\frac{4j\pi}{\tilde{N}_k}
\right)\right)-i\sum_{k=1}^M\rho(z_k)\left(\sum_{j=0}^{\tilde{N}_k-1}r_k^2\sin\left(2\theta_k+\frac{4j\pi}
{\tilde{N}_k}\right)\right)
\\
\hphantom{\tilde{\mu}_{0,2}}{}
=\sum_{k=1}^M\rho(z_k)\left(\sum_{j=0}^{\tilde{N}_k-1}r_k^2\cos(2\theta_k)\cos\left(\frac{4j\pi}{\tilde{N}
_k}\right)-r_k^2\sin(2\theta_k)\sin\left(\frac{4j\pi}{\tilde{N}_k}\right)\right)
\\
\hphantom{\tilde{\mu}_{0,2}=}{}
-i\sum_{k=1}^M\rho(z_k)\left(\sum_{j=0}^{\tilde{N}_k-1}r_k^2\sin(2\theta_k)\cos\left(\frac{4j\pi}
{\tilde{N}_k}\right)+r_k^2\cos(2\theta_k)\sin\left(\frac{4j\pi}{\tilde{N}_k}\right)\right)
\\
\hphantom{\tilde{\mu}_{0,2}}{}
=\sum_{k=1}^M\rho(z_k)\left(r_k^2\cos(2\theta_k)\big(\sum_{j=0}^{\tilde{N}_k-1}\cos\left(\frac{4j\pi}
{\tilde{N}_k}\right)\big)-r_k^2\sin(2\theta_k)\left(\sum_{j=0}^{\tilde{N}_k-1}\sin\left(\frac{4j\pi}{\tilde{N}
_k}\right)\right)\right)
\\
\hphantom{\tilde{\mu}_{0,2}=}{}
-i\sum_{k=1}^M\rho(z_k)\left(r_k^2\sin(2\theta_k)\left(\sum_{j=0}^{\tilde{N}_k-1}\cos\left(\frac{4j\pi}
{\tilde{N}_k}\right)\right)+r_k^2\cos(2\theta_k)\left(\sum_{j=0}^{\tilde{N}_k-1}\sin\left(\frac{4j\pi}{\tilde{N}
_k}\right)\right)\right).
\end{gather*}

It can be shown that
\begin{gather*}
\sum_{j=0}^{\tilde{N}_k-1}\cos\left(\frac{4j\pi}{\tilde{N}_k}\right)=0,
\qquad
\;\sum_{j=0}^{\tilde{N}_k-1}\sin\left(\frac{4j\pi}{\tilde{N}_k}\right)=0,
\qquad
\;\forall\, \tilde{N}_k>2.
\end{gather*}
Then $ \tilde{\mu}_{0,2} = 0$, hence $ \frac{|\tilde{\mu}_{0,2}|}{\tilde{\mu}_{1,1}} = 0$.
\end{proof}

\looseness=-1
One can give a~statistical interpretation of our proposed shape elongation
measure $ \frac{|\tilde{\mu}_{0,2}|}{\tilde{\mu}_{1,1}} $.
Indeed after renormalizing the total ink $ \mu_{0,0} = \sum\limits_{k=1}^N \rho(z_k) $ to one, one can view the pixel
intensities as describing a~discrete probability distribution.
The standard deviation matrix of that distribution is then determined by the third row of the Pascal
triangle, as stated in the following lemma:
\begin{lemma}
Consider the discrete image $ I $ as a~bivariate distribution with the joint probability mass function $ P(X =
x_k-x_0, Y = y_k - y_0) = \frac{\rho(z_k)}{\mu_{0,0}}$, $k = 1,2, \dots, N$.
The covariance matrix $ \Sigma $ of that distribution is given by
\begin{gather*}
\Sigma=
\begin{pmatrix}
\dfrac{\tilde{\mu}_{1,1}+\text{\rm Re}(\tilde{\mu}_{0,2})}{2\mu_{0,0}}&-\dfrac{\text{\rm Im}(\tilde{\mu}_{0,2})}
{2\mu_{0,0}}
\vspace{1mm}\\
-\dfrac{\text{\rm Im}(\tilde{\mu}_{0,2})}{2\mu_{0,0}}&\dfrac{\tilde{\mu}_{1,1}-\text{\rm Re}(\tilde{\mu}_{0,2})}
{2\mu_{0,0}}
\end{pmatrix}
.
\end{gather*}
\end{lemma}

\begin{proof}
Observe that
\begin{gather*}
\frac{\tilde{\mu}_{1,1}}{\mu_{0,0}}=\sum_{k=1}^N(z_k-z_0)(\bar{z}_k-\bar{z}_0)\frac{\rho(z_k)}{\mu_{0,0}}
=\sum_{k=1}^N\big((x_k-x_0)^2+(y_k-y_0)^2\big)\frac{\rho(z_k)}{\mu_{0,0}},
\\
\frac{\tilde{\mu}_{0,2}}{\mu_{0,0}}=\sum_{k=1}^N(\bar{z}_k-\bar{z}_0)^2\frac{\rho(z_k)}{\mu_{0,0}}
\\
\hphantom{\frac{\tilde{\mu}_{0,2}}{\mu_{0,0}}}{}
=\sum_{k=1}^N\big((x_k-x_0)^2-(y_k-y_0)^2\big)\frac{\rho(z_k)}{\mu_{0,0}}-2i\sum_{k=1}
^N(x_k-x_0)(y_k-y_0)\frac{\rho(z_k)}{\mu_{0,0}}.
\end{gather*}
Write $ \Sigma =
\begin{pmatrix}
\sigma_x^2 & \rho_{XY}\sigma_x\sigma_y
\\
\rho_{XY}\sigma_x\sigma_y & \sigma_y^2
\end{pmatrix}
$.  Since the random variables $X$, $Y $  both have zero mean, we have
\begin{gather*}
\sigma_x^2=\sum_{k=1}^N(x_k-x_0)^2\frac{\rho(z_k)}{\mu_{0,0}},
\qquad
\sigma_y^2=\sum_{k=1}^N(y_k-y_0)^2\frac{\rho(z_k)}{\mu_{0,0}},
\\
\rho_{XY}\sigma_x\sigma_y=\sum_{k=1}^N(x_k-x_0)(y_k-y_0)\frac{\rho(z_k)}{\mu_{0,0}}.
\end{gather*}

Then one can check that
\begin{gather*}
\frac{\tilde{\mu}_{1,1}}{2\mu_{0,0}}+\frac{\text{Re}(\tilde{\mu}_{0,2})}{2\mu_{0,0}}=\sigma_x^2,
\qquad
-\frac{\text{Im}(\tilde{\mu}_{0,2})}{2\mu_{0,0}}=\rho_{XY}\sigma_x\sigma_y,
\qquad
\frac{\tilde{\mu}_{1,1}}{2\mu_{0,0}}-\frac{\text{Re}(\tilde{\mu}_{0,2})}{2\mu_{0,0}}=\sigma_y^2. \tag*{\qed}
\end{gather*}
  \renewcommand{\qed}{}
\end{proof}

Recall that, in order to obtain the standard deviation $m_2(\theta) $  of the projection of a~bivariate
distribution onto the line with direction vector $(x, y)=r(\cos\theta, \sin\theta)$,  one can simply project
the standard deviation matrix $\Sigma $  onto $(x, y): \;
\begin{pmatrix}
\cos\theta&\sin\theta
\end{pmatrix}
\Sigma
\begin{pmatrix}
\cos\theta
\\
\sin\theta
\end{pmatrix}
=m_2(\theta)$.

It is easy to check that the relationship between the shape elongation
 $ \frac{|\tilde{\mu}_{0,2}|}{\tilde{\mu}_{1,1}}$ and the eigenva\-lues~$\lambda_{\max}$,~$\lambda_{\min}$ of the
standard deviation matrix $\Sigma $  is
 $ \frac{|\tilde{\mu}_{0,2}|}{\tilde{\mu}_{1,1}}=|\frac{\lambda_{\max}-\lambda_{\min}}{\lambda_{\max}+\lambda_{\min}}|$.

\subsection{Rotational symmetry}

We have seen in the last section that an image $I $  having an
 $ \tilde{N} $-FRS has $\tilde{\mu}_{0,2}=0$.
More generally, we have the following lemmas, which was used in~\cite{Andrew} as the basis for a~HAZMAT
sign recognition method.
\begin{lemma}
Let $\tilde{N}$ be a~$($finite$)$ integer.
If an image $I $  has an $\tilde{N} $-fold rotation symmetry and if $\frac{l-j}{\tilde{N}}$ is not an integer,
then $\tilde{\mu}_{j, l}=0$.  Conversely, if $\tilde{\mu}_{j, l}=0 $  for all $\frac{l-j}{\tilde{N}}$ that are
not an integer, then the image $I $  has an $\tilde{N} $-fold rotation symmetry\footnote{An analogue for the  ``if''  part of this lemma for the case of a~continuous image can be found
in~\cite{Moment}.}.
\end{lemma}
\begin{proof}
 $ \Rightarrow$ Let us rotate $ I $ clockwise around the origin by $ \frac{2\pi}{\tilde{N}} $.
Due to its symmetry, the rotated image $ I' $ must be the same as the original one.
In particular, it must hold
\begin{gather*}
\tilde{\mu}_{j, l}'=e^{2\pi i(l-j)/\tilde{N}}\tilde{\mu}_{j, l}=\tilde{\mu}_{j, l}.
\end{gather*}
Since $ \frac{l-j}{\tilde{N}} $ is not an integer, this equation can be fulf\/illed only if $ \tilde{\mu}_{j, l} =0$.

 $ \Leftarrow\; $ Suppose $ \tilde{\mu}_{j,0} = 0 $ for any $ \frac{j}{\tilde{N}} $ not an integer of some f\/inite
integer $ \tilde{N} $.
Let us rotate $ I $ clockwise around the origin by $ \frac{2k\pi}{\tilde{N}} $ for each $ k =
1,2, \dots, \tilde{N}-1$, then $ \tilde{\mu}_{j,0}' = e^{-2\pi ijk/\tilde{N}}\tilde{\mu}_{j,0} $ for each $ k $ and
all $ j = 0,1, \dots, N-1$.  For $\frac{jk}{\tilde{N}}\in \mathbb{Z}$, it is easy to check that $ e^{-2\pi
ijk/\tilde{N}} = 1$, hence $ \tilde{\mu}_{j,0}' = \tilde{\mu}_{j,0}$.
For $\frac{jk}{\tilde{N}}\notin \mathbb{Z}$, $\frac{j}{\tilde{N}} $ is not an integer either.
Then $ \tilde{\mu}_{j,0} = 0 = \tilde{\mu}_{j,0}' $.
In this way we have $ \tilde{\mu}_{j,0}' = \tilde{\mu}_{j,0} $ for all $ j = 0,1, \dots, N-1 $.
Since in the proof of Lemma~\ref{mutorholemma}, we showed
that $\{\tilde{\mu}_{j,0}\}_{j=0}^{N-1} $ uniquely determine $ I$, then we can conclude
that $ I $ and $ I' $ are the same for any rotation with angle $ \frac{2k\pi}{\tilde{N}}$,
$k = 1,2, \dots, \tilde{N}-1 $.
Therefore the image~$ I $ has an~$ \tilde{N} $-fold rotation symmetry.
\end{proof}

\begin{remark}
As $ N $ increases, more and more columns of the Pascal triangle $ T^r(I) $ become zero, so that in the limit
case, as $ N \rightarrow \infty$, all entries $ \mu_{j, l} $ with $ j\neq l $ of $ T^r(I)$, for any order $ r$, vanish.
This limit case corresponds to $ \infty $-fold rotation symmetry (e.g.~circles), which however does not occur among discrete images.
\end{remark}

\subsection{Ref\/lection symmetry}

Consider the ref\/lection of an image about the line through the origin
with direction vector $  %section $  $
\begin{pmatrix}
\cos(\theta) & \sin(\theta)
\end{pmatrix}
^T$, $\theta\in(-\frac{\pi}{2}, \frac{\pi}{2}]$.  The point $ z_k = x_k + iy_k $ is mapped to $ \underline{z}_k =
\bar{z}_ke^{i2\theta} =  (x_k\cos(2\theta) + y_k\sin(2\theta) ) + i (x_k\sin(2\theta) -
y_k\cos(2\theta) ) $ under the ref\/lection with its pixel intensity $ \rho(z_k) $ staying the same.
Then the new complex moment is
\begin{gather*}
\underline{\mu}_{j, l}=\sum_{k=1}^N\underline{z}_k^j\bar{\underline{z}}_k^l\rho(\underline{z}_k)=\sum_{k=1}
^N\big(\bar{z}_ke^{i2\theta}\big)^j\big(z_ke^{-i2\theta}\big)^l\rho(z_k)
\\
\hphantom{\underline{\mu}_{j, l}}{}
=\sum_{k=1}^N e^{i2(j-l)\theta}\bar{z}_k^jz_k^l\rho(z_k)=e^{i2(j-l)\theta}\sum_{k=1}^N\bar{z}
_k^jz_k^l\rho(z_k)
=\mu_{l, j}e^{i2(j-l)\theta},
\qquad
\forall\,  j, l\in\mathbb{Z}_{\geq0}.
\end{gather*}

Therefore the moment matrix for the new image $ \underline{I} $ after ref\/lection is
\begin{gather*}
\tau_N(\underline{I})=
\begin{pmatrix}
1&0&0&\cdots&0
\\
0&e^{-i2\theta}&0&\cdots&0
\\
0&0&e^{-i4\theta}&\cdots&0
\\
\vdots&\vdots&\vdots&\ddots&\vdots
\\
0&\cdots&\cdots&0&e^{-i2(N-1)\theta}
\end{pmatrix}
\tau_N(I)^{T}
\begin{pmatrix}
1&0&0&\cdots&0
\\
0&e^{i2\theta}&0&\cdots&0
\\
0&0&e^{i4\theta}&\cdots&0
\\
\vdots&\vdots&\vdots&\ddots&\vdots
\\
0&\cdots&\cdots&0&e^{i2(N-1)\theta}
\end{pmatrix}
.
\end{gather*}

We can conclude from the above relation that if an image is symmetric with respect to the $x$-axis (i.e.~$ \theta = 0 $), then we will have $ \tau_N(I) = \tau_N(I)^T$, i.e.\
 $ \mu_{j, l} = \mu_{l, j} $ for all $ j, l \in \mathbb{Z}_{\geq 0} $.
Since $ \mu_{j, l} = \bar{\mu}_{l, j} $ by def\/inition, this means that all the $ \mu_{j, l}$'s are
real\footnote{For the continuous analogue, the fact that ref\/lecting an object horizontally transforms
complex moments into their conjugate was previously noted in~\cite{Moment}.}.

Similarly, if an image is symmetric with respect to the $y$-axis (i.e.\
 $ \theta = \frac{\pi}{2} $), we can conclude that $ \mu_{j, l}$'s are real for $j$, $l $ of the same parity
and $ \mu_{j, l}$'s are imaginary for $j$, $l $ of opposite parity.

More generally, we have the following result:
\begin{lemma}
\label{L13}
A discrete image is symmetric with respect to reflections about a~line through the origin with direction $
\begin{pmatrix}
\cos\theta_0 & \sin\theta_0
\end{pmatrix}
^T $
\begin{gather*}
\Longleftrightarrow
\quad
\tan(l-j)\theta_0=-\frac{\text{\rm Im}(\mu_{j, l})}{\text{\rm Re}(\mu_{j, l})},
\qquad
j, l=0,1, \dots.
\end{gather*}
\end{lemma}
\begin{proof}
Notice that with ref\/lection symmetry, we have
\begin{gather*}
m_n(\theta_0-\theta)=\frac{1}{2^n}\sum_{k=1}^N\big(z_ke^{-i(\theta_0-\theta)}+\bar{z}_ke^{i(\theta_0-\theta)}
\big)^n\rho(z_k)
\\
\hphantom{m_n(\theta_0-\theta)}{}
=\frac{1}{2^n}\sum_{k=1}^N\big(\bar{z}_ke^{i2\theta_0}e^{-i(\theta_0+\theta)}+z_ke^{-i2\theta_0}
e^{i(\theta_0+\theta)}\big)^n\rho(z_k)
\\
\hphantom{m_n(\theta_0-\theta)}{}
=\frac{1}{2^n}\sum_{k=1}^N\big(z_k'e^{-i(\theta_0+\theta)}+\bar{z}_k'e^{i(\theta_0+\theta)}\big)^n\rho\big(\bar{z}
_k'e^{i2\theta_0}\big)
\qquad
\big(\text{denote}~z_k'=\bar{z}_ke^{i2\theta_0}\big)
\\
\hphantom{m_n(\theta_0-\theta)}{}
=\frac{1}{2^n}\sum_{k=1}^N\big(z_k'e^{-i(\theta_0+\theta)}+\bar{z}_k'e^{i(\theta_0+\theta)}\big)^n\rho(z_k')
=m_n(\theta_0+\theta).
\end{gather*}

Conversely, if $m_n(\theta_0-\theta)=m_n(\theta_0+\theta) $  for an image
 $ I= \{ (z_k, \rho(z_k) ) \}_{k=1}^N$,  then for its ref\/lection image
 $ I_r= \{ (\bar{z}_ke^{i2\theta_0}, \rho(z_k) ) \}_{k=1}^N$,  there is
 $ m_n^r(\theta_0+\theta)=m_n(\theta_0-\theta)=m_n(\theta_0+\theta) $  for any $\theta\in(-\pi, \pi]$.  From
Lemmas~\ref{mutorholemma} and~\ref{mtoTn} we can conclude that the image reconstructed from
 $ \big\{m_n^r(\theta_j), \,  j=1, \ldots, n+1\big\}_{n=0}^{N-1}$ and
 $ \big\{m_n(\theta_j), \, j=1, \ldots, n+1\big\}_{n=0}^{N-1}$ will be the same, i.e.\
 $ I_r = I$.  Hence the image has ref\/lection symmetry.

By equation~\eqref{mnmukl}, we know
\begin{gather*}
m_n(\theta_0+\theta)=\frac{1}{2^n}\sum_{l=0}^n
\begin{pmatrix}
n
\\
l
\end{pmatrix}
\mu_{l, n-l}e^{i(n-2l)(\theta_0+\theta)}, \\
m_n(\theta_0-\theta)=\frac{1}{2^n}\sum_{l=0}^n
\begin{pmatrix}
n
\\
l
\end{pmatrix}
\mu_{l, n-l}e^{i(n-2l)(\theta_0-\theta)}.
\end{gather*}
Then $ m_n(\theta_0+\theta) = m_n(\theta_0-\theta)$, i.e.\
 $ m_n(\theta_0+\theta) - m_n(\theta_0-\theta) = 0 $ can be written as
\begin{gather*}
\frac{1}{2^n}\sum_{l=0}^n
\begin{pmatrix}
n
\\
l
\end{pmatrix}
\mu_{l, n-l}e^{i(n-2l)\theta_0}\big(e^{i(n-2l)\theta }-e^{-i(n-2l)\theta }\big)=0
\\
\Leftrightarrow
\quad
\sum_{l=0}^n
\begin{pmatrix}
n
\\
l
\end{pmatrix}
\mu_{l, n-l}e^{i(n-2l)\theta_0}\big(2i\sin(n-2l)\theta\big)=0
\\
\Leftrightarrow
\quad
\sum_{l=0}^n
\begin{pmatrix}
n
\\
l
\end{pmatrix}
\mu_{l, n-l}e^{i(n-2l)\theta_0}\sin(n-2l)\theta=0
\\
\Leftrightarrow
\quad
\sum_{l=0}^{\lfloor\frac{n-1}{2}\rfloor}
\begin{pmatrix}
n
\\
l
\end{pmatrix}
\sin(n-2l)\theta\big(\mu_{l, n-l}e^{i(n-2l)\theta_0}-\mu_{n-l, l}e^{i(2l-n)\theta_0}\big)=0
\\
\Leftrightarrow
\quad
\sum_{l=0}^{\lfloor\frac{n-1}{2}\rfloor}
\begin{pmatrix}
n
\\
l
\end{pmatrix}
\sin(n-2l)\theta2i\,\text{Im}\big(\mu_{l, n-l}e^{i(n-2l)\theta_0}\big)=0
\\
\Leftrightarrow
\quad
\sum_{l=0}^{\lfloor\frac{n-1}{2}\rfloor}
\begin{pmatrix}
n
\\
l
\end{pmatrix}
\sin(n-2l)\theta\big(\text{Re}(\mu_{l, n-l})\sin(n-2l)\theta_0+\text{Im}(\mu_{l, n-l}
)\cos(n-2l)\theta_0\big)=0.
\end{gather*}
The above equation is true for any $ \theta\in \big({-}\frac{\pi}{2}, \frac{\pi}{2}\big]$, so it is true if and only if
\begin{gather*}
\text{Re}(\mu_{l, n-l})\sin(n-2l)\theta_0+\text{Im}(\mu_{l, n-l})\cos(n-2l)\theta_0=0 ,
\qquad
l=0,1, \dots, \left\lfloor\frac{n-1}{2}\right\rfloor,
\end{gather*}
i.e.
\begin{gather}
\label{theta0mujl}
\tan(n-2l)\theta_0=-\frac{\text{Im}(\mu_{l, n-l})}{\text{Re}(\mu_{l, n-l})},
\qquad
l=0,1, \dots, \left\lfloor\frac{n-1}{2}\right\rfloor.
\end{gather}
Since $ \mu_{n-l, l} = \bar{\mu}_{l, n-l}$, \eqref{theta0mujl} can be written as
\begin{gather}
\label{theta0mujl2}
\tan(l-j)\theta_0=-\frac{\text{Im}(\mu_{j, l})}{\text{Re}(\mu_{j, l})},
\qquad
\forall\,  j=0,1, \dots, \quad l=0,1, \dots.\hspace{30mm} \qed
\end{gather}
\renewcommand{\qed}{}
\end{proof}

In this way, for a~given discrete image, we can check~\eqref{theta0mujl2} to decide whether it is symmetric
with respect to a~certain line or not.
If it is, then the line is at an angle $ \theta $ given by
\begin{gather*}%\label{symaxis}
\theta=
\begin{cases}
\arctan\left(-\dfrac{\text{Im}(\mu_{j, j+1})}{\text{Re}(\mu_{j, j+1})}\right)&\text{if $ \text{Re}(\mu_{j, j+1}
)\neq0$, $\forall\,  j=0,1, \dots, $ }
\\
\dfrac{\pi}{2}&\text{otherwise.}
\end{cases}
\end{gather*}

\section{Experiments and results}

To illustrate the application of the Pascal triangle to symmetry
detection, we used our results to design a~simple ref\/lection symmetry detection method\footnote{The reader
interested in the application of the Pascal triangle to the detection of rotational symmetries is invited
to read~\cite{Andrew}.}.
We tested this method on images from the MPEG-7  CE Shape-1 Part-B data set\footnote{Shape data for the MPEG-7 core experiment
CE-Shape-1, \url{http://www.cis.temple.edu/~latecki/TestData/mpeg7shapeB.tar.gz}.}. %~\cite{MPEG-7}.
The data set includes 1400 binary images.
The images are divided into 70 object classes, each object class containing 20 images.
All our ground truth data and classif\/ication results can be downloaded
from \url{https://engineering.purdue.edu/~mboutin/symmetric_shapes}.

\subsection{Horizontally symmetric object detection experiment}

\looseness=-1
In this experiment, we identif\/ied images
that have a~horizontal axis of ref\/lection symmetry using only the f\/irst four rows of the Pascal triangle.
Our data set consists of 320 shapes from the MPEG-7  shape database.
Specif\/ically, we included all 20 images contained in each of the
following 16 classes:  Bird, Device1-Device5, Device7-Device9, Watch, Cup, Dog, Flatf\/ish, Glas, Hat, Tree.

\begin{figure}[t]\centering \setcounter{subfigure}{0}
\subfigure[Set 1a shapes classified as symmetric.]{\centering
\includegraphics[width=3cm]{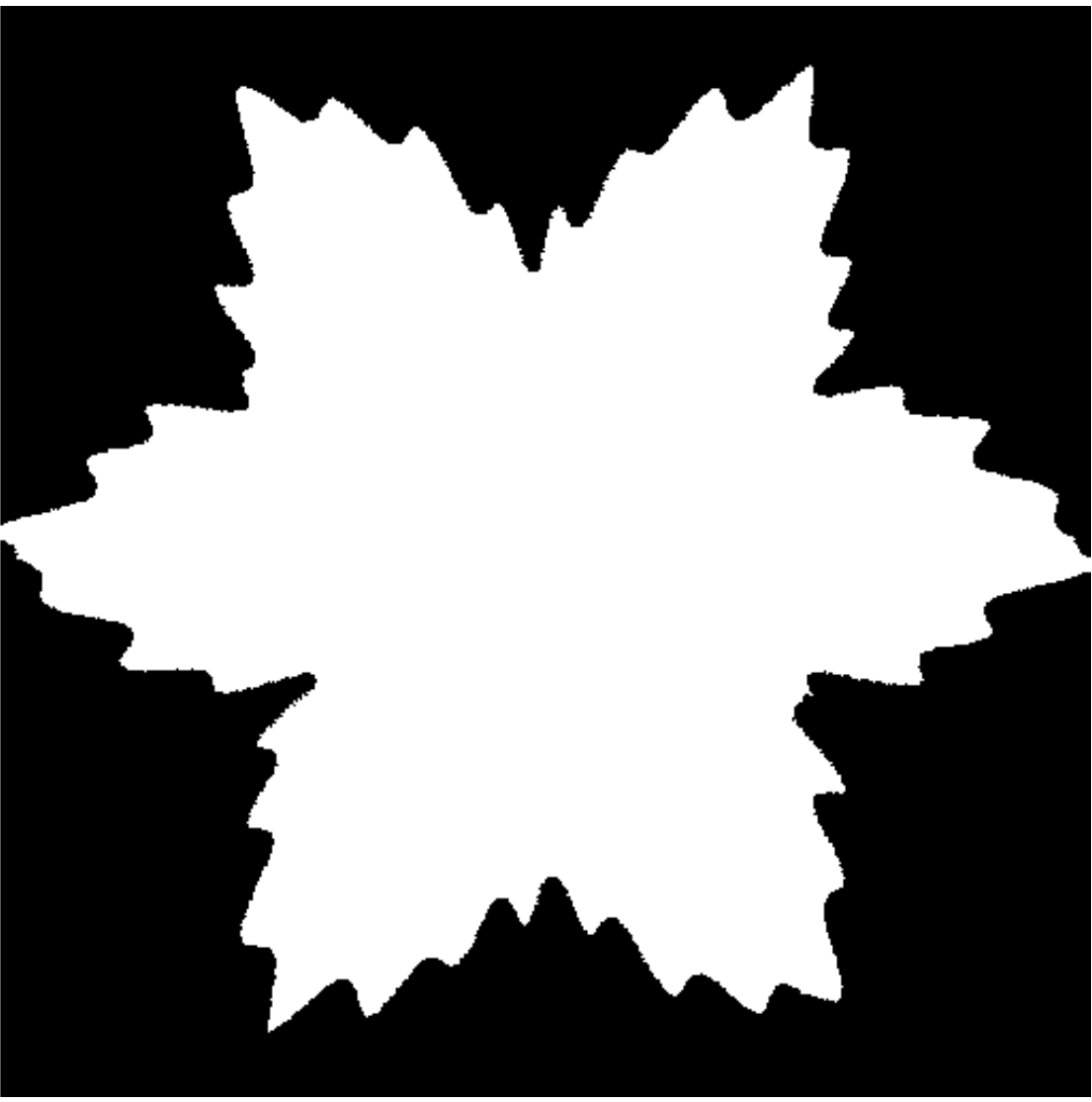} \quad
\includegraphics[width=3cm]{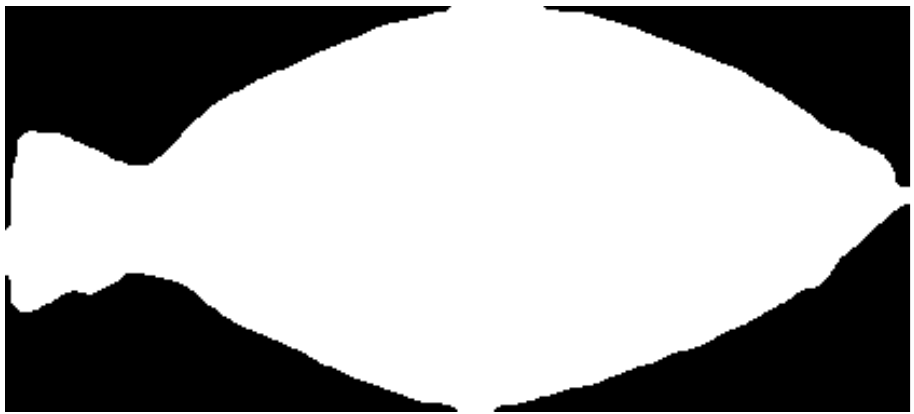}
}
\qquad \subfigure[Set 1a shapes classified as not symmetric.]{\centering
\includegraphics[width=3cm]{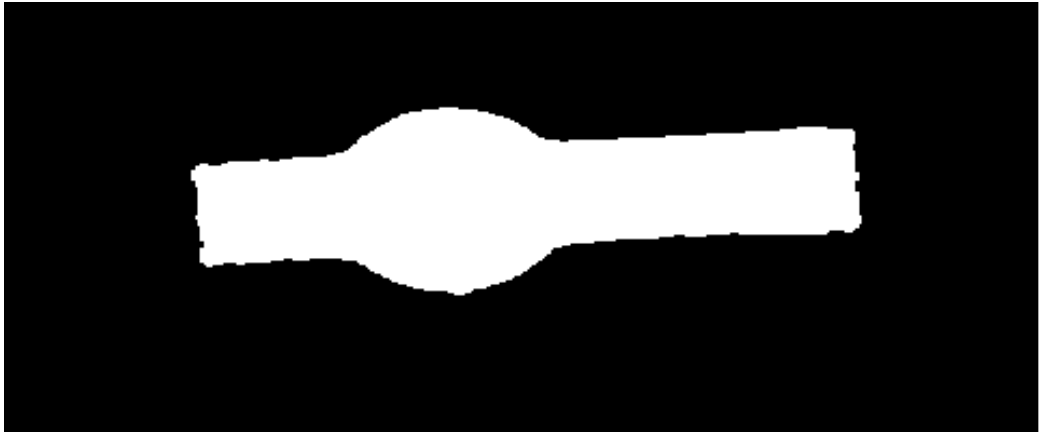} \quad
\includegraphics[width=3cm]{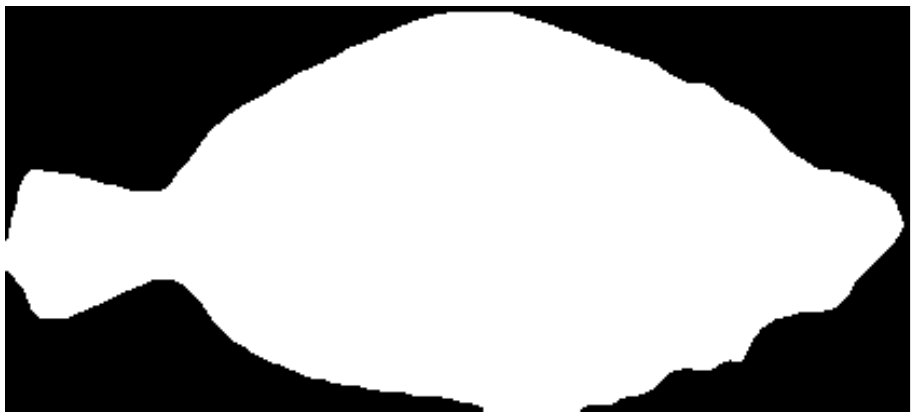}
}\\
\subfigure[Set 2a shapes (not in Set 1a) classified as symmetric.]{\centering
\includegraphics[width=3cm]{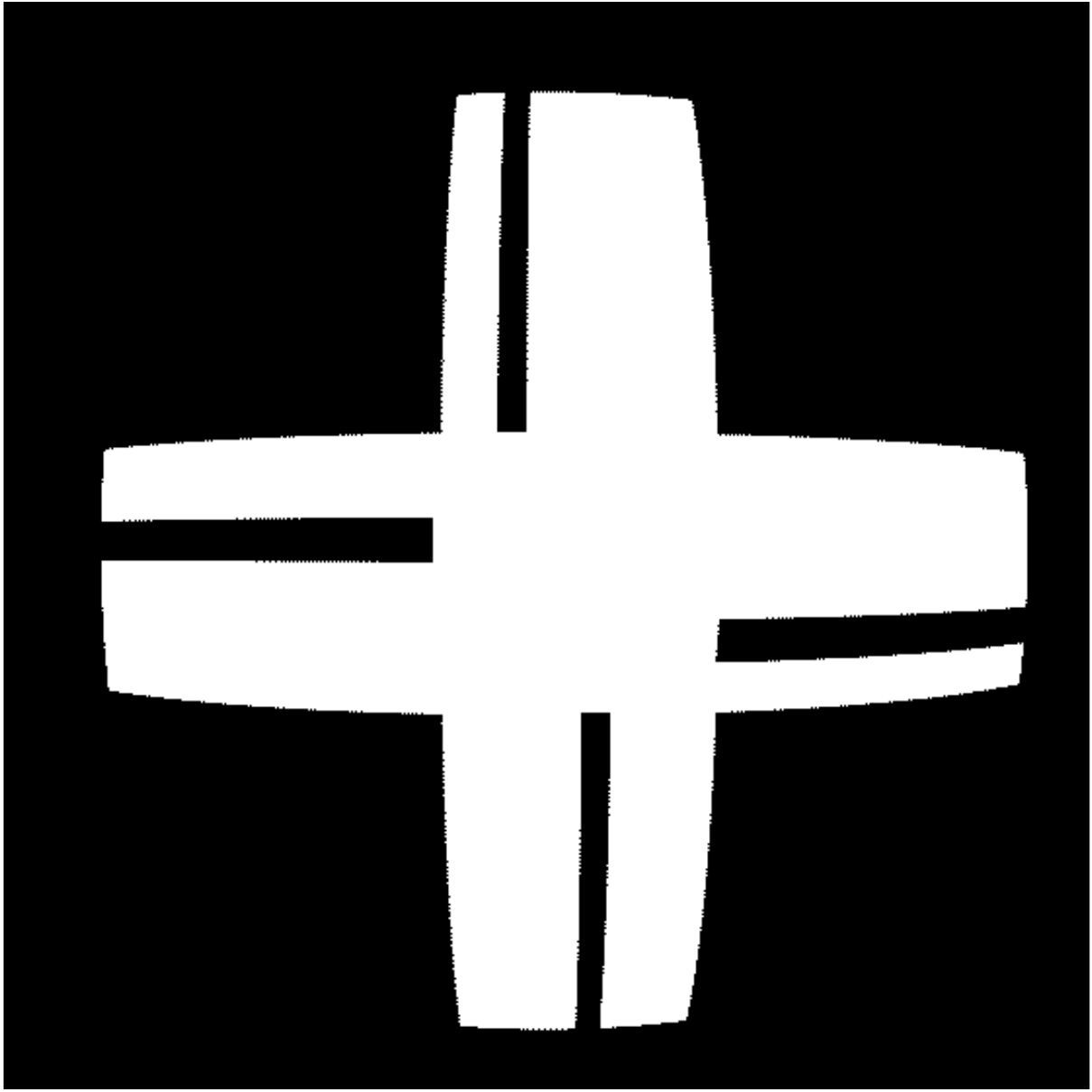} \quad
\includegraphics[width=3cm]{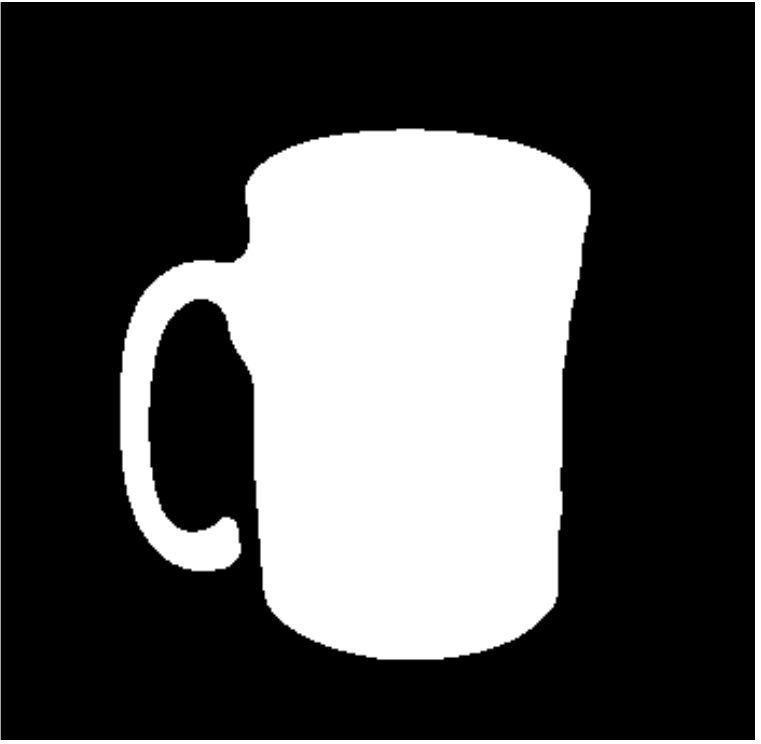}
}
\qquad \subfigure[Set 2a shapes (not in Set 1a) classified as not symmetric.]{\centering
\includegraphics[width=3cm]{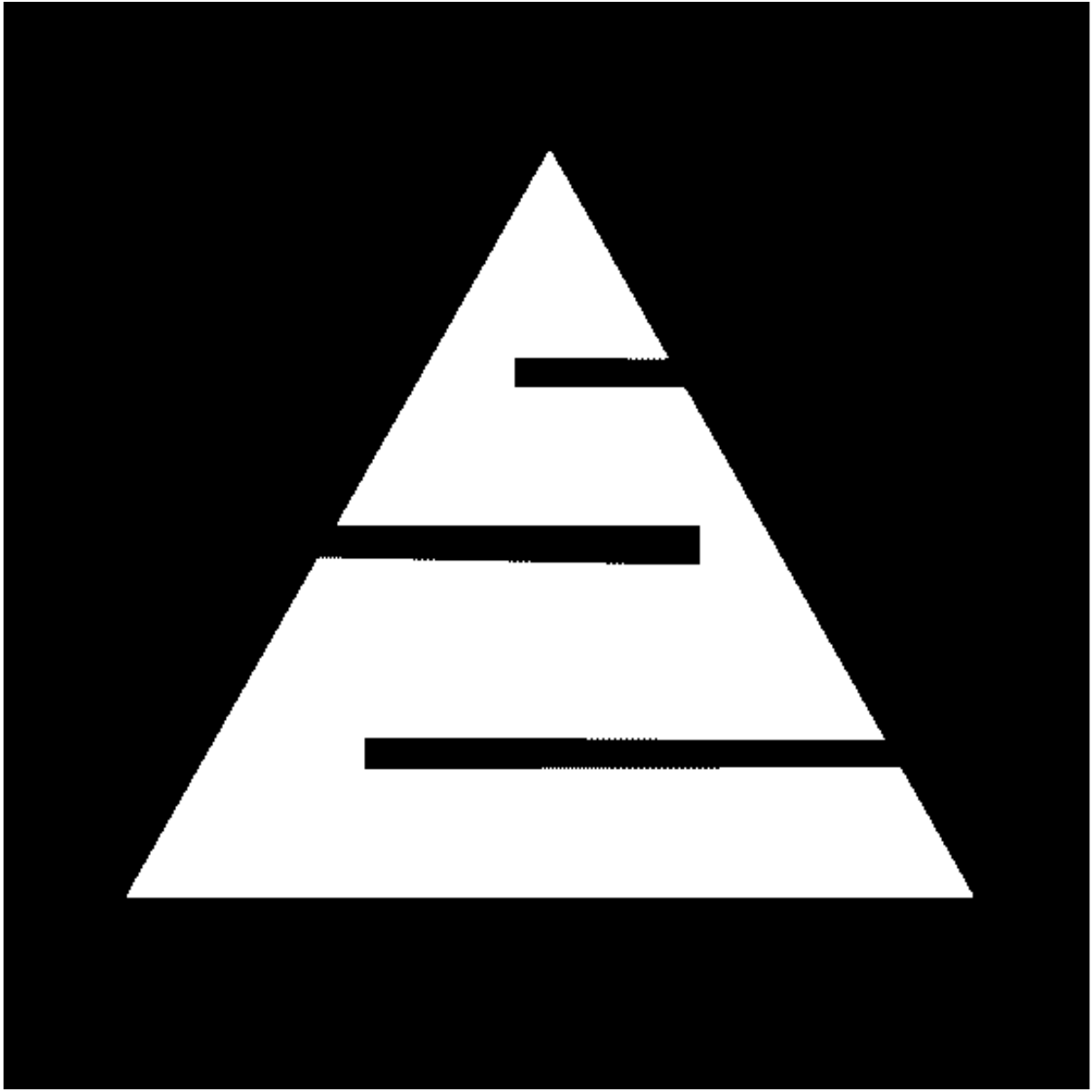} \quad
\includegraphics[width=3cm]{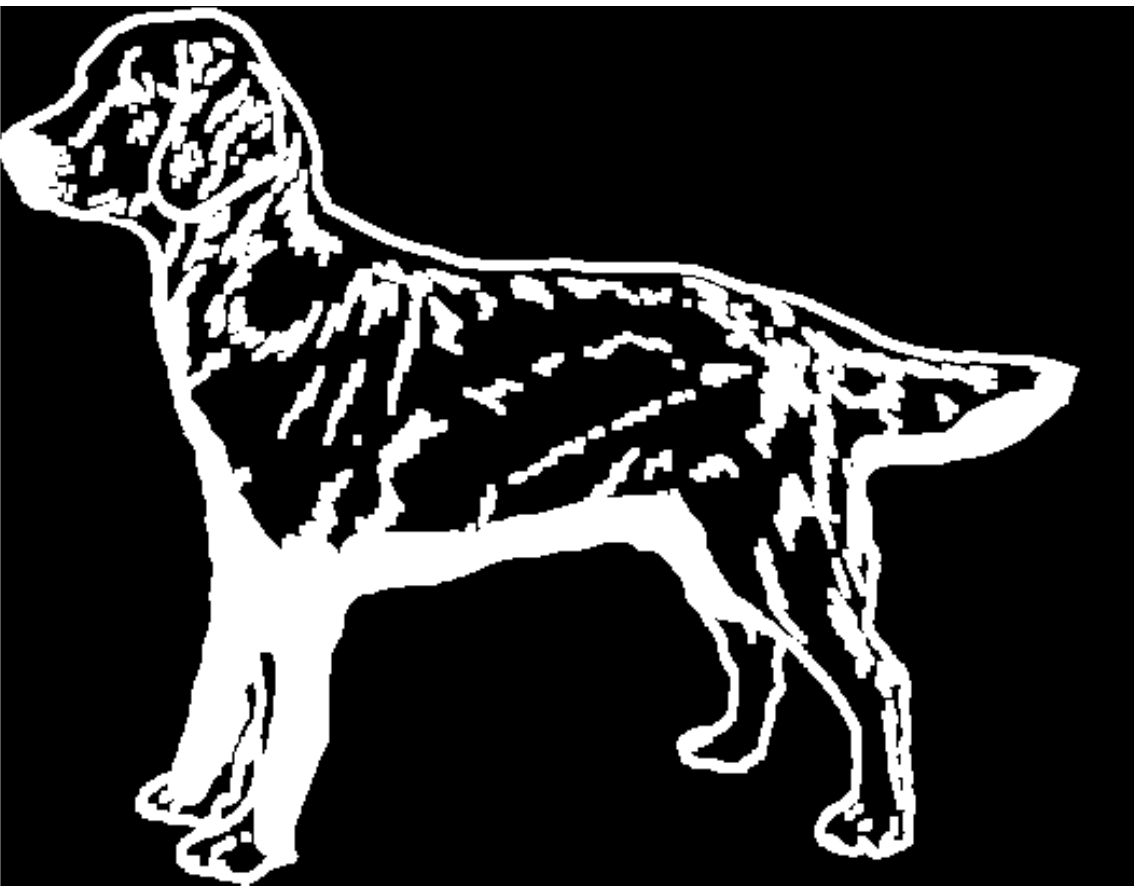}
}
\caption{Small selection of shapes chosen from the MPEG-7  shape database, $r = 0.07$. }\label{samplesp}
\end{figure}

We f\/irst manually divided the data into two sets.
One set, called Set~1a, was assigned all objects that appeared to have a~clear horizontal symmetry axis, up to
some minor details.
The remaining set, called Set~2a, was assigned the remaining images.
Set~1a and Set~2a contain~113 and~207 images respectively.
As one can see by inspecting Set~2a (see for example the images in Fig.~\ref{samplesp}(c) and (d)), our
classif\/ication was quite strict.
Indeed, we excluded many objects that could be declared symmetric under a~greater tolerance for error.
We thus created a~second grouping allowing for more errors: Set~1b and Set~2b, which are the data sets resulting from
this more lenient def\/inition of symmetry.

Recall that, by Lemma~\ref{L13}, an image has a~horizontal axis of symmetry if and only if all the entries of
its Pascal triangle are real.
Since we are focusing on the symmetry of the object contained in the image, as opposed to the image
itself, we need to consider the translation invariant Pascal triangle consisting of the centralized
moments $ \tilde{\mu}_{j, l}$.
Thus horizontally symmetric objects should be recognizable by considering the magnitude of the imaginary
part of each of its centralized moments.
Note that $ \tilde{\mu}_{0,0} $ and $ \tilde{\mu}_{0,1} $ are always real and that $ \tilde{\mu}_{2,0} =
\bar{\tilde{\mu}}_{0,2} $.
Thus, if we restrict ourselves to the f\/irst four rows of the Pascal triangle, for simplicity, then
horizontally symmetric objects are characterized by the fact that $ \tilde{\mu}_{0,2}$,
$\tilde{\mu}_{0,3} $ and $ \tilde{\mu}_{1,2} $ are real.
In other words, objects that are approximately symmetric should have $ \tilde{\mu}_{0,2}$,
$\tilde{\mu}_{0,3} $ and $ \tilde{\mu}_{1,2} $ with an imaginary part close to zero.
In order to remove the scale ambiguity resulting from the arbitrary scale used to describe the pixel
coordinates, we followed the approach described in Section~\ref{sec4}, Example~\ref{exp2} to invariantize our
coordinates with respect to scaling.
Our specif\/ic classif\/ication criteria were:
\begin{gather*}
\text{if}
\qquad
\text{Im}\left(\frac{\tilde{\mu}_{0,2}}{\tilde{\mu}_{1,1}}\right)^2+\text{Im}\left(\frac{\tilde{\mu}_{0,3}}
{\tilde{\mu}_{1,1}^{3/2}}\right)^2+\text{Im}\left(\frac{\tilde{\mu}_{1,2}}{\tilde{\mu}_{1,1}^{3/2}}\right)^2<r^2 ,
\\
\qquad \text{then object is symmetric,}
\\
\text{else}
\\
\qquad \text{object is not symmetric,}
\end{gather*}
where $ r $ is a~variable threshold.

In our experiments, we varied the threshold $ r $ from $ 0.005 $ to $ 0.15$.
For each value of $ r$, we classif\/ied every image as either ``symmetric'' or not symmetric using the above
mentioned criteria.
We also computed the precision, recall, and accuracy for each value of $ r$, where
\begin{gather*}
\text{precision}=\frac{\text{number of true positives}}{\text{number of true positives}
+\text{false positives}},
\\
\text{recall}=\frac{\text{number of true positives}}{\text{number of true positives}+\text{false negatives}
},
\\
\text{accuracy}=\frac{\text{number of true positives}+\text{true negatives}}
{\text{number of true positives}+\text{true negatives}+\text{false positives}+\text{false negatives}}
.
\end{gather*}

\begin{figure}[t]\centering\setcounter{subfigure}{0}
\subfigure[Detection of horizontally symmetric objects using Set~1a and Set~2a data set and the first four rows of the Pascal triangle. The max accuracy is $ 83.75\%$.]{
\includegraphics[width=11.0cm]{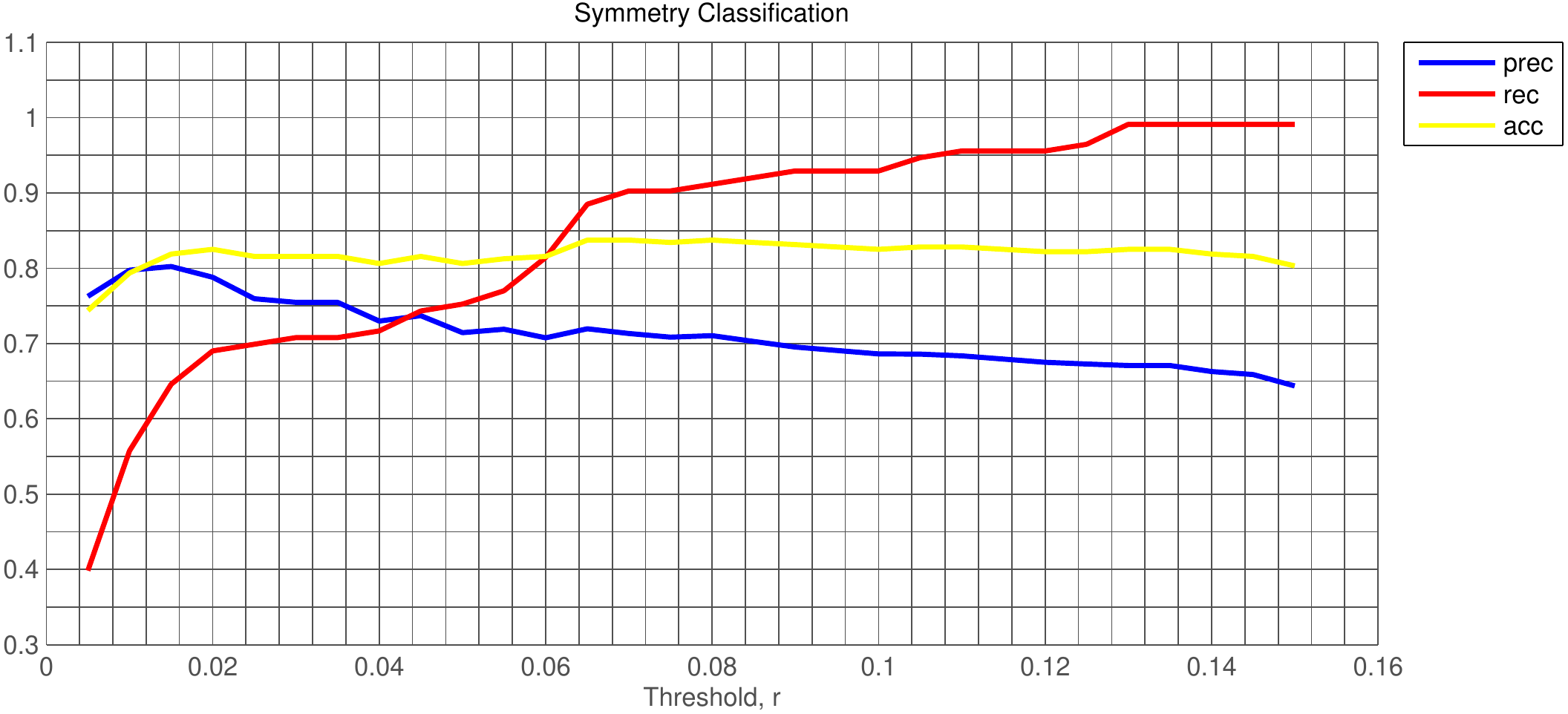}
}
\\
\subfigure[Detection of horizontally symmetric objects using Set~1b and Set~2b data set and the first four rows of the Pascal triangle. The max accuracy is $ 96.25\%$.]{
\includegraphics[width=11.0cm]{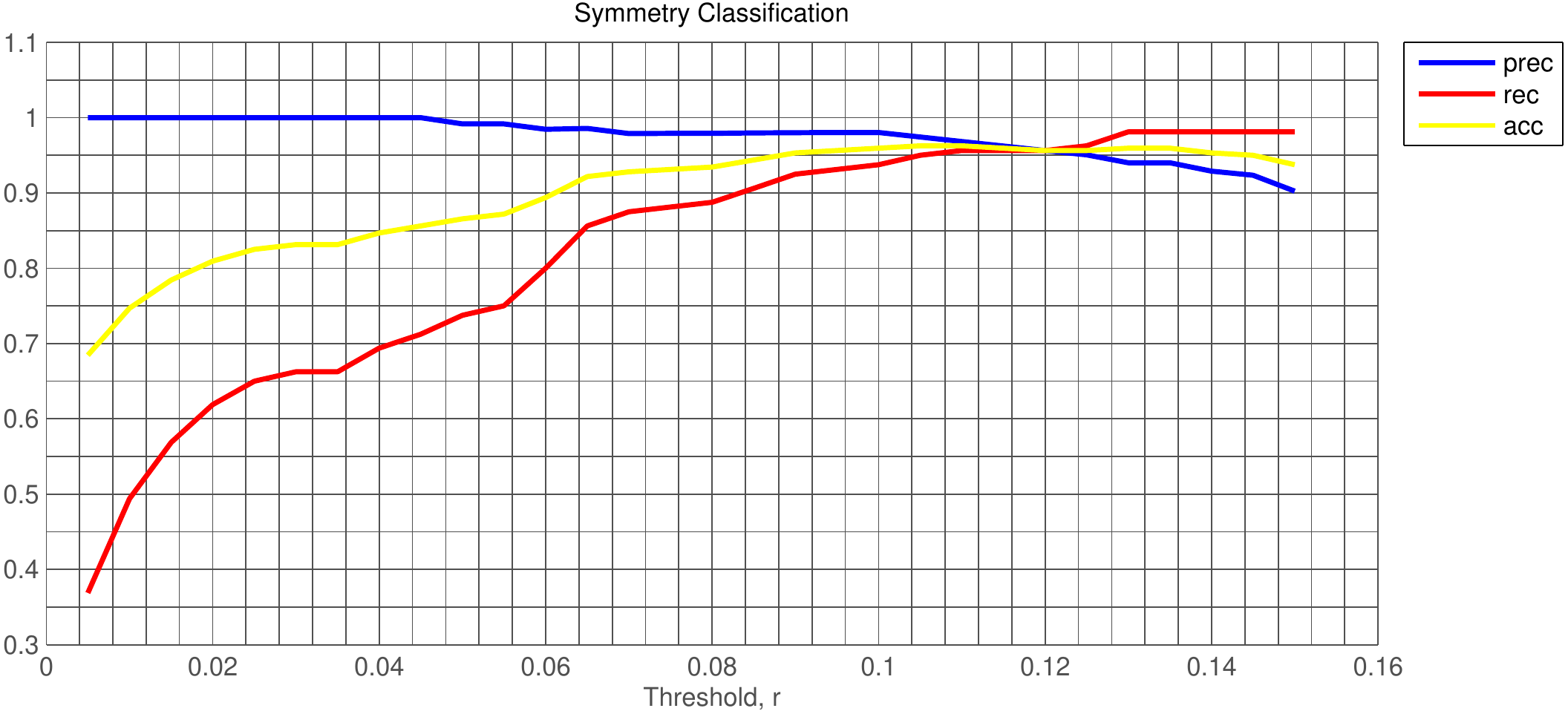}
}
\caption{Precision, recall and accuracy for $r $  from $0.005 $  to $0.15 $  at increments of $0.005$.  }\label{prec}
\end{figure}

The results obtained when using the data sets Set~1a and Set~2a are plotted in Fig.~\ref{prec}(a), and those
obtained using Set~1b and Set~2b are plotted in Fig.~\ref{prec}(b).
Observe that the maximum accuracy for the f\/irst data set, $83.75\%$  (obtained around $r = 0.07 $), goes up to $96.25\% $  (obtained around $r = 0.11 $) for the second data set. This is consistent with the fact that the second data set was constructed using a greater tolerance for error:  after all, we are only using the f\/irst four rows of the triangle to classify the shape. Indeed, the shapes that were moved from Set~2a to Set~1b caused the number of false positive to decrease and thus the precision to increase correspondingly. Fig.~\ref{samplesp} illustrates some of our results.

\subsection{Symmetric objects (any axis) detection experiment} In this experiment we identif\/ied images
that have an axis of ref\/lection symmetry.
Our data set consists of 200 shapes from the following 10 classes of the MPEG-7  shape database:  Beetle,
Bell, Bird, Butterf\/ly, Camel, Cattle, Classic, Crown, Horseshoe, Lizzard.

We f\/irst manually divided the data set into two sets.
One set, called Set~1, was assigned all objects in the classes of Beetle, Bell, Butterf\/ly, Crown and
Horseshoe.
Because these objects all represent shapes that have a~natural axis of symmetry.
The remaining set, called Set~2, was assigned the images in the classes of Bird, Camel, Cattle, Classic and
Lizzard.
Set~1 and Set~2 each contains 100 images.

Recall that, by Lemma~\ref{L13}, an image is symmetric with respect to ref\/lections about a~line through
the origin with direction  $
\begin{pmatrix}
\cos\theta_0 & \sin\theta_0
\end{pmatrix}
^T $ if and only if $ \tan(l-j)\theta_0 = -\frac{\text{Im}(\mu_{j, l})}{\text{Re}(\mu_{j, l})}$, for all $ j, l \in
\mathbb{Z}_+$.
Since we are focusing on the symmetry of the object contained in the image, as opposed to the image
itself, we need to consider the translation invariant Pascal triangle consisting of the centralized
moments $ \tilde{\mu}_{j, l}$.
Thus the symmetry axis of symmetric objects should be recognizable by considering the arc tangent of the
negative ratio of the imaginary part and the real part of each of its centralized moments.
Note that the arc tangent of an angle are always between $-\frac{\pi}{2} $ and $ \frac{\pi}{2}$, hence there will
be some ambiguity when deciding $ \theta_0 $ from $ \arctan(k\theta_0) $ with $ |k|>1$.
Also note that $ \tilde{\mu}_{0,1} $ is always zero and $ \tilde{\mu}_{j, l} = \bar{\tilde{\mu}}_{l, j} $.
Thus for simplicity, we characterized the symmetry axis of symmetric objects by the fact that
\begin{gather*}
\theta_0=\arctan\left(-\frac{\text{Im}(\tilde{\mu}_{1,2})}{\text{Re}(\tilde{\mu}_{1,2})}
\right)=\arctan\left(-\frac{\text{Im}(\tilde{\mu}_{2,3})}{\text{Re}(\tilde{\mu}_{2,3})}\right)=\arctan\left(-\frac{\text{Im}
(\tilde{\mu}_{3,4})}{\text{Re}(\tilde{\mu}_{3,4})}\right).
\end{gather*}
In other words, objects that are approximately symmetric with respect to an axis of angle~$ \theta_0 $ should
have $ \arctan\big({-}\frac{\text{Im}(\tilde{\mu}_{1,2})}{\text{Re}(\tilde{\mu}_{1,2})}\big)$,
$\arctan\big({-}\frac{\text{Im}(\tilde{\mu}_{2,3})}{\text{Re}(\tilde{\mu}_{2,3})}\big) $ and $ \arctan\big({-}\frac{\text{Im}(\tilde{\mu}_{3,4})}{\text{Re}(\tilde{\mu}_{3,4})}\big) $ close to each other.
Since we consider the ratio of the imaginary part and real part of each moment, it is not necessary to
remove the scale ambiguity resulting from the arbitrary scale used to describe the pixel coordinates in
this experiment, as the quantities we consider are already invariant under scaling.
Also note that, if the symmetry axis of an image is of angle $ \theta_0 = \frac{\pi}{2}$, then $ \theta_0 =
-\frac{\pi}{2} $ also def\/ines the same symmetry axis, although the arc tangent of the angles near these two
values are quite dif\/ferent.
Taking these into account, our specif\/ic classif\/ication criteria were:
\begin{gather*}
\text{if}
\qquad
\left| |\theta_1|-\frac{\pi}{2}\right| <T,
\qquad
\left| |\theta_2|-\frac{\pi}{2}\right| <T,
\qquad
\left|
|\theta_3|-\frac{\pi}{2}\right| <T,
\\
\qquad \text{then object is symmetric vertically,}
\\
\text{else if}
\qquad
|\theta_1-\theta_2|<T,
\qquad
|\theta_2-\theta_3|<T,
\qquad
|\theta_3-\theta_1|<T,
\\
\qquad \text{then object is symmetric with symmetry axis}~\theta_0=\frac{\theta_1+\theta_2+\theta_3}{3},
\\
\text{else}
\\
\qquad \text{object is not symmetric,}
\end{gather*}
where $ \theta_1 = \arctan\big({-}\frac{\text{Im}(\tilde{\mu}_{1,2})}{\text{Re}(\tilde{\mu}_{1,2})}\big)$, $\theta_2 =
\arctan\big({-}\frac{\text{Im}(\tilde{\mu}_{2,3})}{\text{Re}(\tilde{\mu}_{2,3})}\big)$, $\theta_3 =
\arctan\big({-}\frac{\text{Im}(\tilde{\mu}_{3,4})}{\text{Re}(\tilde{\mu}_{3,4})}\big) $ and $ T $ is a~variable threshold.

In our experiments, we varied the threshold~$T $  from $1^\circ $  to~$15^\circ$.
For each value of~$T$,  we classif\/ied every image as either  ``symmetric''  or not symmetric using the above
mentioned criteria.
And for each symmetric image, we found its symmetry axis.
We also computed the precision, recall, and accuracy for each value of~$T $.

\begin{figure}[t]\centering
  \includegraphics[width=11.0cm]{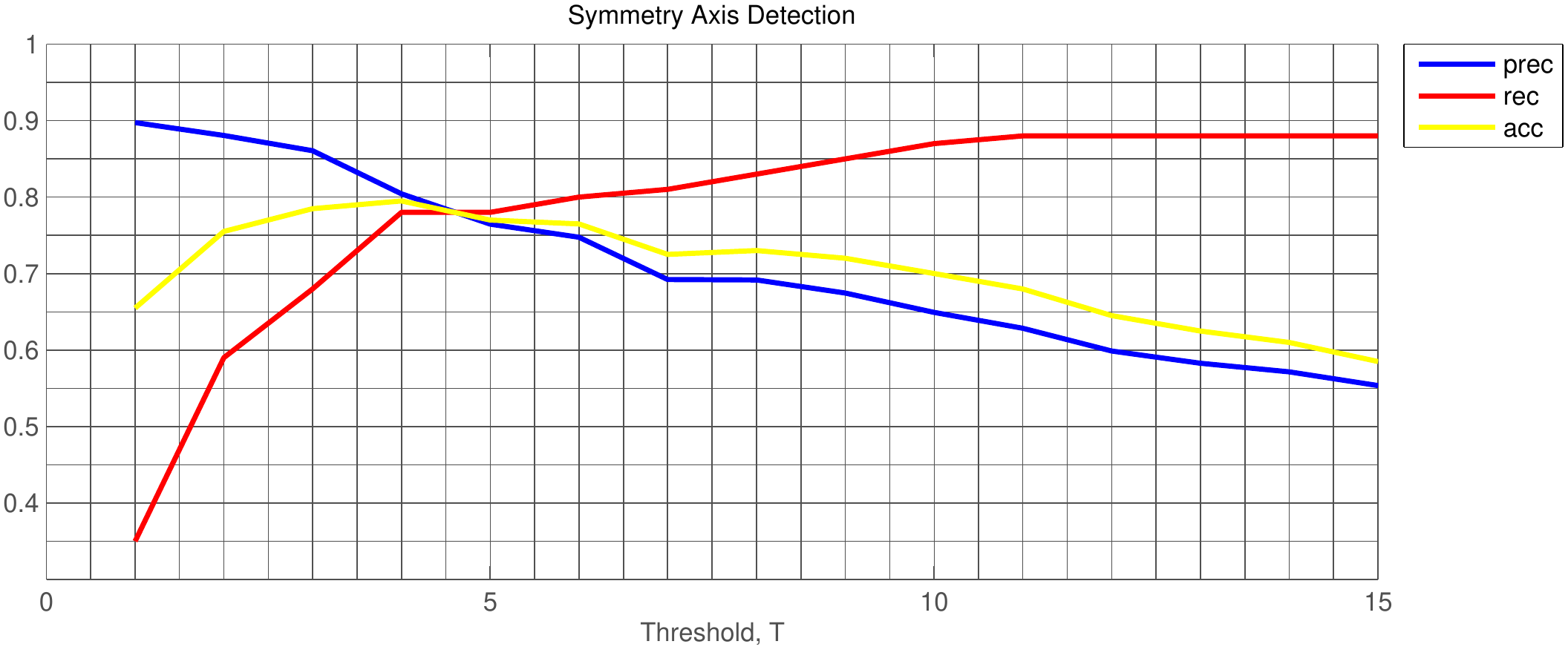}
  \caption{Symmetric objects detection using $\tilde{\mu}_{2,2}$, $\tilde{\mu}_{2,3}$ and $\tilde{\mu}_{3,4}$.  The max accuracy is $79.5\%$. }\label{accSAdetection}
\end{figure}

\begin{figure}[t] \centering\setcounter{subfigure}{0}
\subfigure[Set 1 shapes classified as symmetric with axis $53.0^{\circ}$ and $94.5^{\circ}$ respectively.]{\centering
\includegraphics[width=3.0cm]{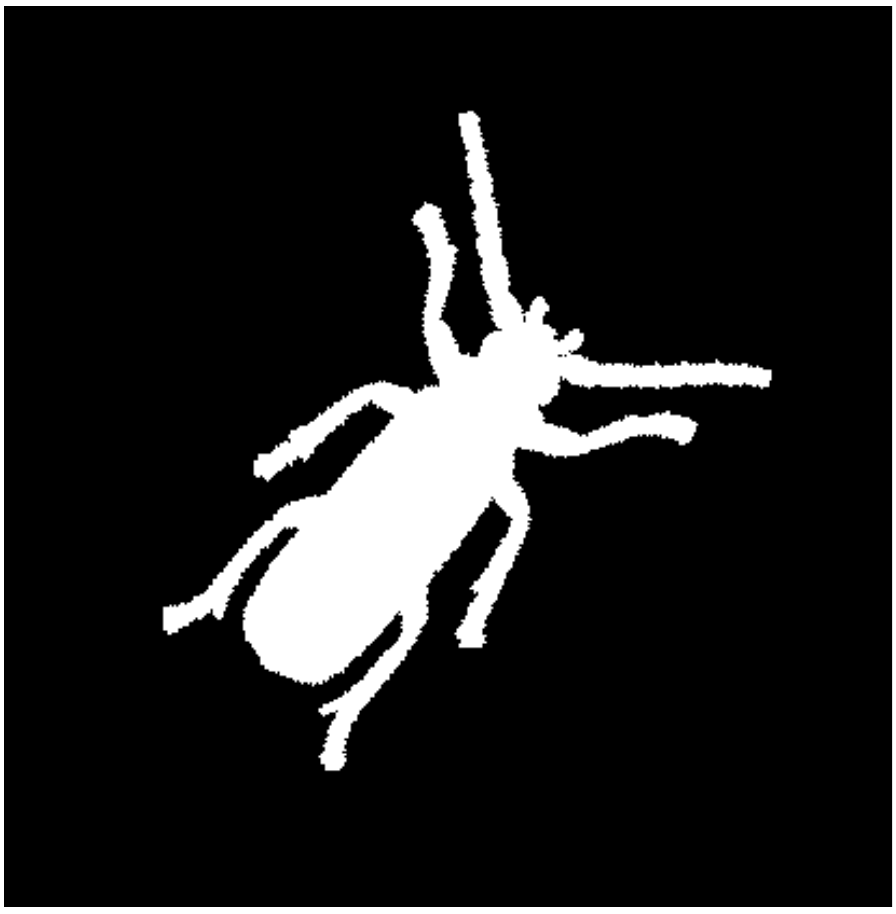} \quad
\includegraphics[width=3.0cm]{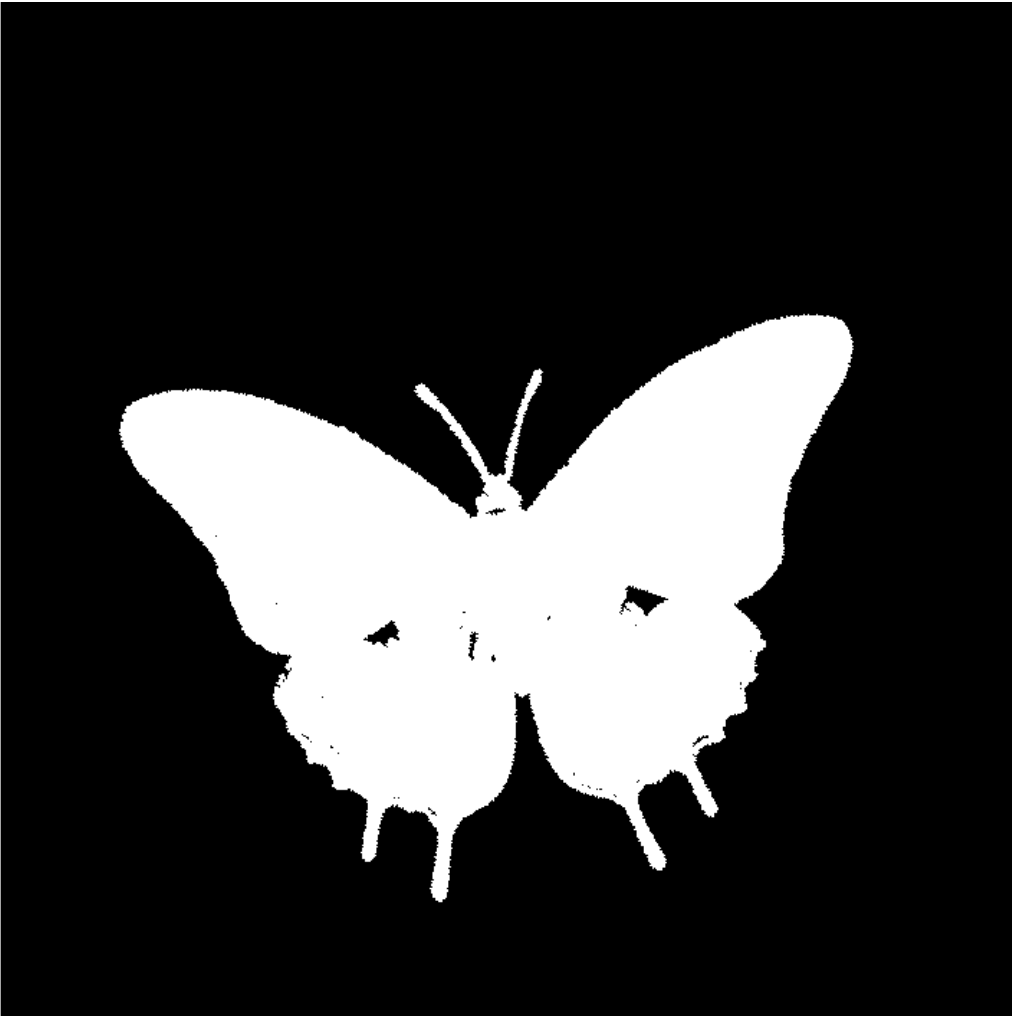}
}
\qquad \subfigure[Set 1 shapes classified as not symmetric.]{\centering
\includegraphics[width=3.0cm]{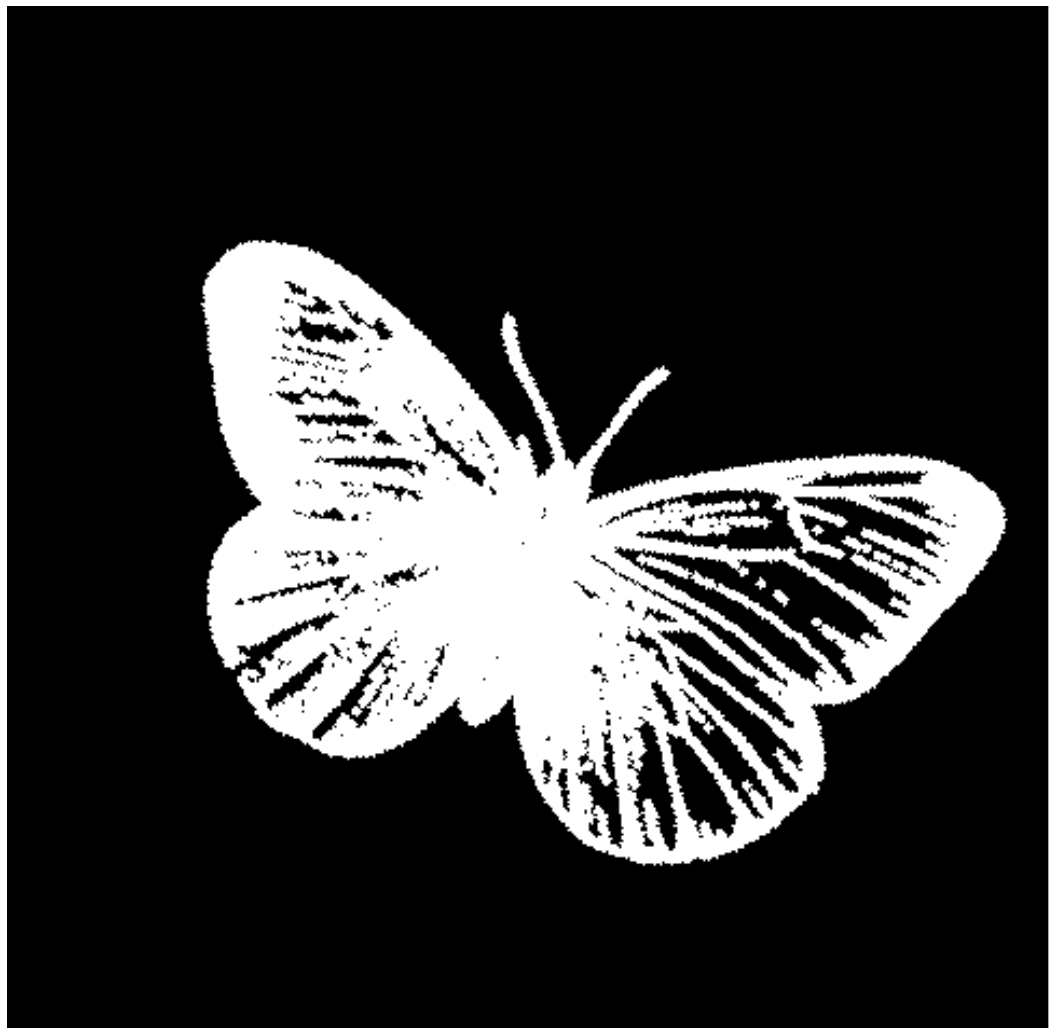} \quad
\includegraphics[width=3.0cm]{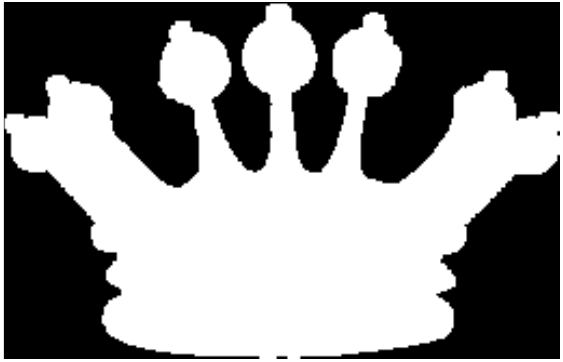}
}\\[-2pt]
\subfigure[Set 2 shapes classified as symmetric with axis $99.2^{\circ}$ and $136.1^{\circ}$ respectively.]{\centering
\includegraphics[width=3.0cm]{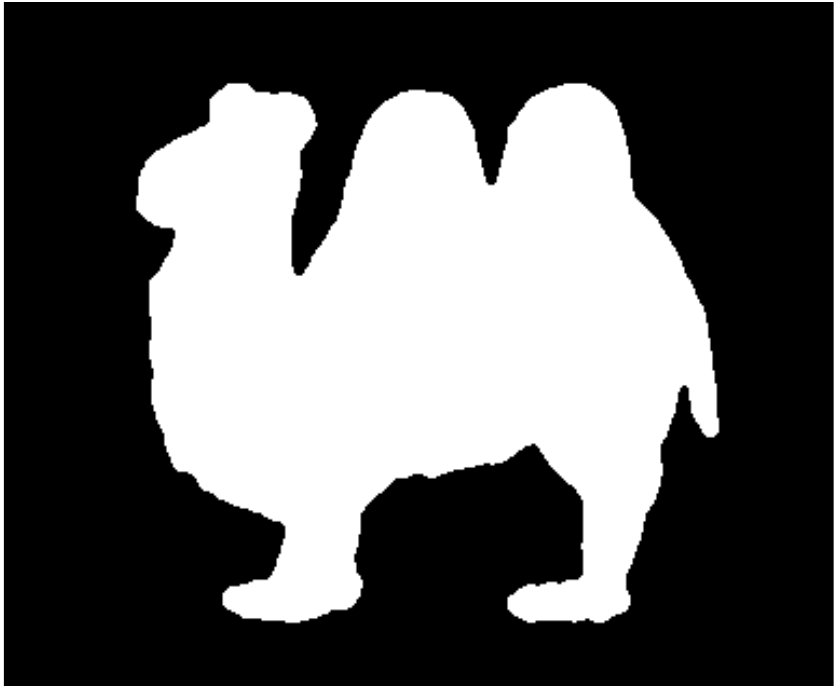} \quad
\includegraphics[width=3.0cm]{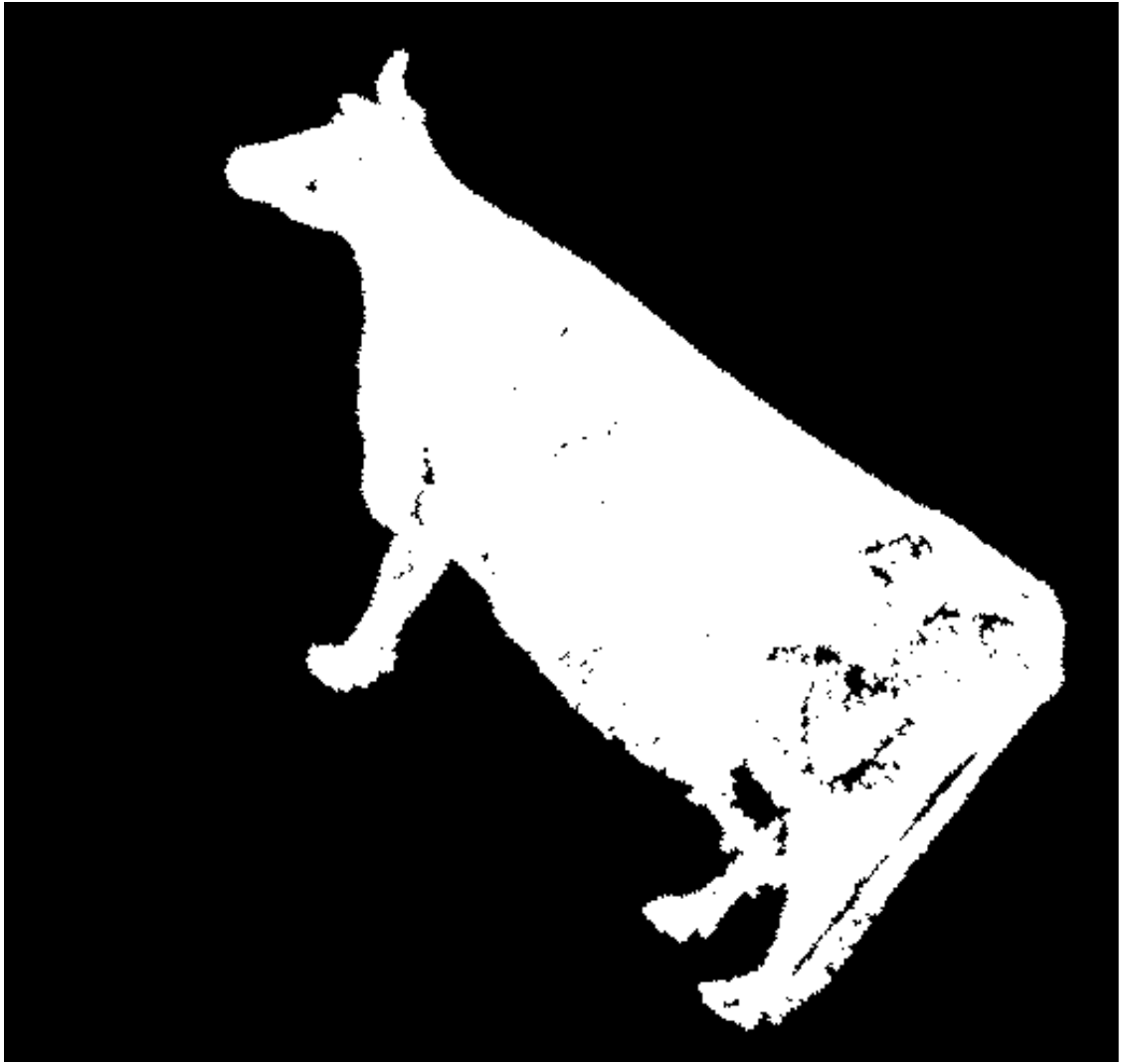}
}
\qquad \subfigure[Set 2 shapes classified as not symmetric.]{\centering
\includegraphics[width=3.5cm]{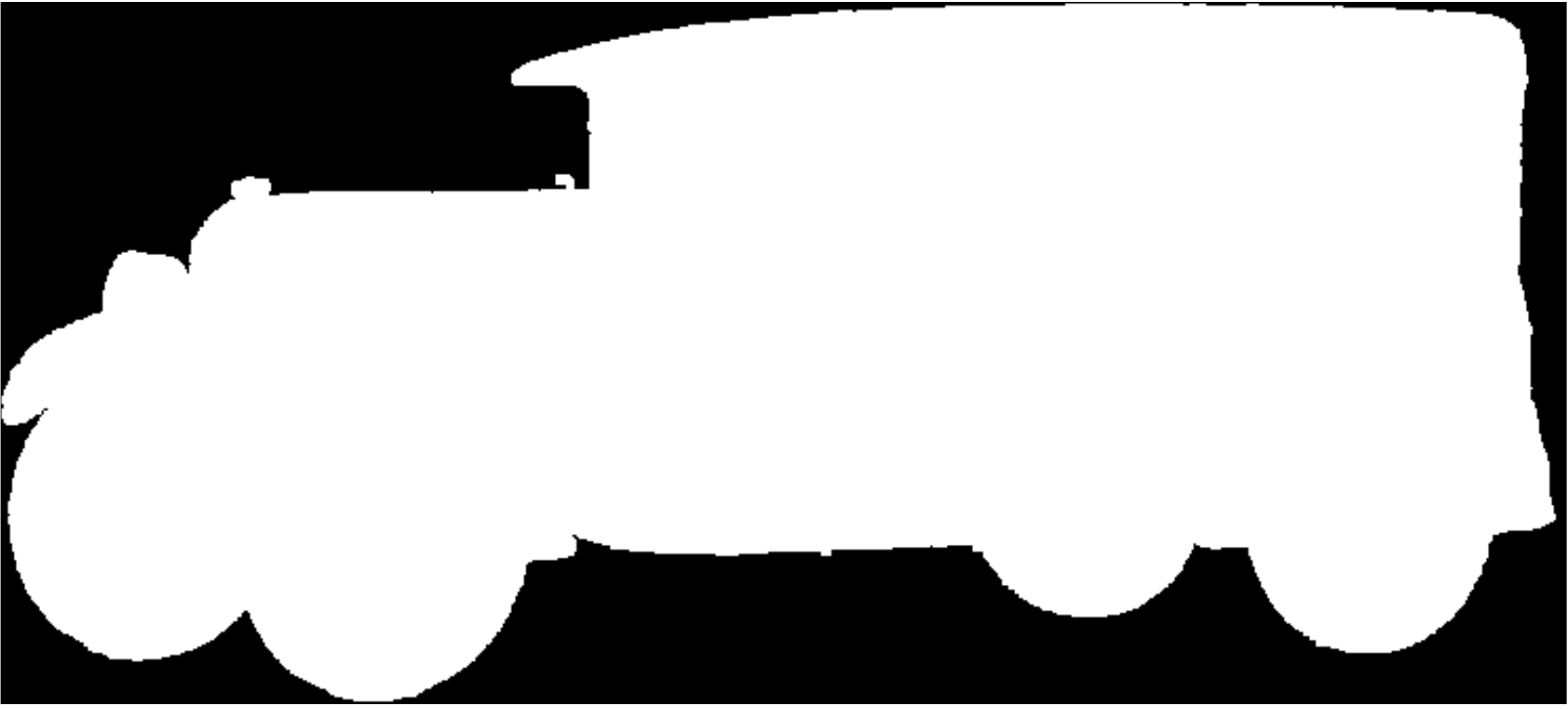} \quad
\includegraphics[width=3.0cm]{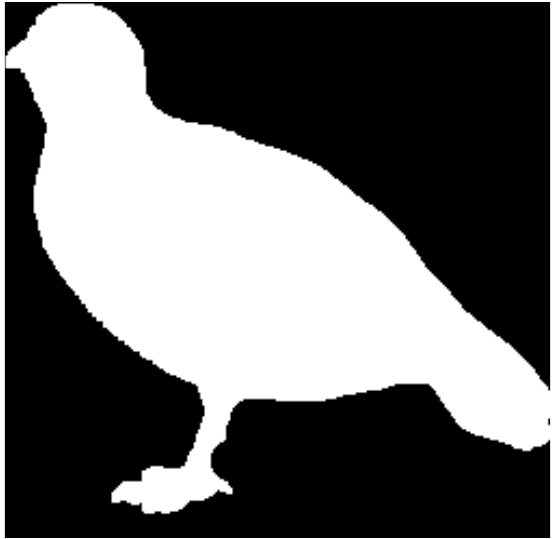}
}
\caption{Small selection of symmetric objects detection results for images in Set~1 and Set~2 with threshold $T=5$.  }\label{symaxisdet}
\vspace{-1.5mm}

\end{figure}

The classif\/ication results obtained when using the data sets Set~1 and Set~2 are plotted in
Fig.~\ref{accSAdetection}.
Observe that the maximum accuracy for the data set is $79.5\%$  at $T=4^\circ $.
The accuracy of the experiment could be improved by using more moments:  after all, we are only using three moments to classify the shapes. Indeed, using more moments would yield a more selective criterion, which should decrease the number of false positives and increase the number of true negatives. Fig.~\ref{symaxisdet} illustrates some of our results.

\section{Conclusion and future work}

We have introduced the Pascal triangle of a~discrete image, which is
constructed using complex-valued moments.
We obtained the relationship between the triangle and the Fourier series coef\/f\/icients of the moment of
the Radon transform of the image, that is, each row $ n $ of the Pascal triangle contains the coef\/f\/icients of
the Fourier series of the $n$-th order moment of the Radon transform of the image.
This relationship gives the moments a~clear geometric interpretation.
For example, $ \frac{\mu_{0,3}}{8} $ of an image is the coef\/f\/icient of $ e^{i3\theta} $ in the third order
moment $ m_3(\theta) $ of the Radon transform of the image, and thus when $ |\mu_{0,3}| $ is large, then the order
three variation of the skewness of the projection is proportionally large.

\looseness=-1
We showed that the image can be fully reconstructed using a~f\/inite number of rows of the triangle.
This fact, which is specif\/ic to discrete (f\/inite) images, allows us to be able to derive necessary and
suf\/f\/icient conditions for the presence of various symmetries.
It also allows us to conclude that the invariantized Pascal triangle separates the orbits of certain group
actions.
Indeed, by using the moving frame method we were able to invariantize the Pascal triangle with respect to
translation, rotation and scaling, and by using the reconstruction property of the invariantized Pascal
triangle, we were able to show the uniqueness of the reconstruction modulo these transformations.

We tested the application of the Pascal triangle to the recognition of symmetric shapes from the
MPEG-7  shape database.
More specif\/ically, we derived a~simple method to detect horizontal symmetries using the f\/irst four rows
of the triangle.
We then tested this method using 16 object classes.
We also derived a~simple method to detect symmetry axes in objects using entries within only the f\/irst
eight rows of the triangle.
We then tested this method using 10 object classes.

\looseness=-1
Extension of our method to other group actions such as af\/f\/ine transforms should be doable using the
moving frame method.
Observe that our def\/inition for $\mu_{j, l}$ naturally extends to vector valued pixel intensities.
Therefore, it should be straightforward to extend our framework to the case of color images.
We are looking forward to also extending this work to the case of~3D objects.

\vspace{-1.5mm}

\subsection*{Acknowledgments}

This research was supported in parts by NSF grant CCF-0728929.
%Address correspondence to Mireille Boutin (mboutin@purdue.edu).

\vspace{-2.5mm}

\pdfbookmark[1]{References}{ref}

\end{document}